\documentclass[journal=jctcce,manuscript=article]{achemso}
\setkeys{acs}{articletitle = true}

\usepackage{graphicx,amsmath,color}
\usepackage[utf8]{inputenc}
\DeclareUnicodeCharacter{2010}{-}

\newcommand{\PQG}{\mathcal{PQG}}
\newcommand{\PQGT}{\mathcal{PQGT}}

\title{Benchmarking the variational reduced density matrix theory in
	the doubly-occupied configuration interaction space with integrable
	pairing models}

\author{A.~Rubio-Garc\'{\i}a}
\email{alvaro.rubiog@iem.cfmac.csic.es}
\affiliation{Instituto de Estructura de la Materia, CSIC, Serrano 123, 28006
	Madrid, Spain}
\author{D. R.~Alcoba}
\affiliation{Departamento de F\'isica, Facultad de Ciencias Exactas y Naturales, Universidad de Buenos Aires,
	Ciudad Universitaria, 1428 Buenos Aires, Argentina}
\altaffiliation{Instituto de F\'isica de Buenos Aires, Consejo Nacional de Investigaciones Cient\'ificas y T\'ecnicas,
	Ciudad Universitaria, 1428 Buenos Aires, Argentina}
\author{P. Capuzzi}
\affiliation{Departamento de F\'isica, Facultad de Ciencias Exactas y Naturales, Universidad de Buenos Aires,
	Ciudad Universitaria, 1428 Buenos Aires, Argentina}
\altaffiliation{Instituto de F\'isica de Buenos Aires, Consejo Nacional de Investigaciones Cient\'ificas y T\'ecnicas,
	Ciudad Universitaria, 1428 Buenos Aires, Argentina}
\author{J.~Dukelsky}
\affiliation{Instituto de Estructura de la Materia, CSIC, Serrano 123, 28006
	Madrid, Spain}

\begin{document}

\begin{abstract}
  The variational reduced density matrix theory has been
  recently applied with great success to models within the truncated
  doubly-occupied configuration interaction space, which corresponds
  to the seniority zero subspace. Conservation of the seniority
  quantum number restricts the Hamiltonians to be based on the SU(2)
  algebra. Among them there is a whole family of exactly solvable
  Richardson-Gaudin pairing Hamiltonians. We benchmark the variational
  theory against two different exactly solvable models, the
  Richardson-Gaudin-Kitaev and the reduced BCS Hamiltonians. We obtain
  exact numerical results for the so-called $\PQGT$
  $N$-representability conditions in both cases for systems that go
  from 10 to 100 particles. However, when random single-particle
  energies as appropriate for small superconducting grains are considered,
  the exactness is lost but still a high accuracy is obtained.
\end{abstract}

\maketitle
\section{\label{sec:intro}Introduction}

One of the main problems in many-body quantum mechanics, which
includes condensed matter, nuclear physics, and quantum chemistry, is
the so-called exponential wall problem \cite{Kohn2000}, namely, the
exponential growth of the dimension of the Hilbert space with the
number of particles composing the studied system.  A complete
diagonalization of the corresponding Hamiltonian in the many-particle
space provides the exact answer but at  a prohibitively expensive
computational cost. Therefore, research efforts have been focused on
the development of approximate methods capturing the relevant degrees
of freedom present in the wavefunction at a feasible computational
cost, i.e., with a polynomial increase.

A great variety of such approximate methods that have been developed
over the years can be broadly classified into approximations that improve
over a reference state and variational theories.  In the former case,
a standard approach is to start from a mean-field reference state and
improve on this by adding perturbative corrections \cite{Shavitt} or
excitations with increasing complexity within Coupled Cluster Theory
\cite{Bartlett, Hagen}. However, these methods break down in strong
correlation regimes where multi-reference approximations are
needed. New variational methods overcoming this issue were developed
in the last decades.  For instance, variational algorithms like
tensor-network-state approaches
\cite{White1992,Niggemann1997,Schollwock2005,Vidal2007},
 variational Monte Carlo methods
\cite{McMillan1965,Ceperley1977,Mezzacapo2009,Changlani2009}, or
stochastic techniques \cite{Suzuki1977,Prokofev1996,Syljuasen2002,Alet2005} can be made, in
principle, as accurate as the exact diagonalizations while
extending its computational limits beyond. Some of these many-body
methods were recently benchmarked in the hydrogen chain \cite{Motta}.

A very different approach to tackle the exponential wall problem that
is applicable to any correlation regime concentrates on the
second-order reduced density matrix (2RDM)
\cite{Husimi1940,Lowdin1955}, while dispensing with the wavefunction
altogether. The 2RDM is a much more compact object than the
wavefunction and it holds all the necessary information to evaluate
the expectation values of one- and two-particle observables of
physical interest. As the energy of any pairwise-interacting system
can be written as an exact but simple linear function of the 2RDM, it
can be used to variationally optimize this matrix at polynomial cost
\cite{Mayer1955}. This optimization should be constrained to the class
of 2RDMs that can be derived from a wavefunction (or an ensemble of
wavefunctions), the so-called $N$-representable 2RDMs
\cite{Tredgold1957,Coleman1963}. Since the complete characterization
of this class of 2RDMs is known to be a quantum Merlin Arthur (QMA)
complete problem \cite{Liu2007}, one has to use an incomplete set of
necessary but not (in general) sufficient constraints on the 2RDM. The
optimization thus finds a lower bound to the exact ground-state energy
and an approximation to the exact ground-state 2RDM.  Such an
approach, known as the variational second-order reduced density matrix
(v2RDM) method has been applied with different degrees of success in
quantum-chemistry problems \cite{Garrod1975,Nakata2001,
  Mazziotti2002,Zhao2004}, nuclear-physics \cite{Rosina,Exact}, and
condensed-matter \cite{Hubbard1,Hubbard2, Hubbard3}.

Recently, the computational efficiency of the v2RDM method has been
substantially improved for systems whose states can be accurately
described in terms of doubly-occupied single-particle states
only. This lies at the heart of the doubly-occupied configuration
interaction (DOCI) method, widely used in quantum chemistry to reduce
the dimension of the configuration interaction Hilbert space. DOCI
corresponds to the subspace of the Hilbert space of seniority zero,
where the seniority quantum number \cite{Talmi} counts the number of
unpaired particles. It has been recognized that the DOCI subspace
captures most of the static correlations, serving as the first rung on
a seniority ladder leading to the exact full configuration interaction
(CI) solution \cite{Bitautas, Alcoba2013,Limacher2013,Alcoba2014}.
The assumptions in DOCI drastically simplify the structure of the 2RDM
\cite{Weinhold1967,Weinhold1967b} and reduce the scaling of the v2RDM
method \cite{Poelmans2015,Head-Marsden2017,Alcoba2018} while,
expectedly, retain most of the correlation.  Several applications of
the v2RDM for seniority nonconserving Hamiltonians were already
implemented and their accuracy tested against exact diagonalizations
for small systems \cite{Poelmans2015Thesis,Poelmans2015,Alcoba2018}.
Here we will take advantage of the seniority-zero nature of DOCI space
that restricts the Hamiltonians to be seniority conserving and
therefore, to be based on the SU(2) algebra. An important class of
SU(2) Hamiltonians are the pairing Hamiltonians, where the fundamental
physics lies in the specific form of the paired states. The quantum
integrable and exactly solvable Richardson-Gaudin pairing models
\cite{Dukelsky2001, Dukelsky2004,CCT} are ideal Hamiltonians to test
the performance of the v2RDM method within the DOCI space. In this
paper, we will benchmark the method for two different integrable
Richardson-Gaudin models: the Richardson-Gaudin-Kitaev model
\cite{Ortiz} describing a chain of spinless fermions with p-wave
pairing, and the constant pairing or reduced BCS Hamiltonian with
uniform \cite{Richardson} and random single-particle energies
\cite{Delft}. We will also explore the behavior of the method for
increasingly large systems addressing its extensivity properties

\section{Theory}

In second quantization, an $N$-particle Hamiltonian with pairwise interactions can be
written as \cite{Jorgensen1981}
\begin{equation}
H=\sum_{ij}t_{ij}c^\dagger_i c_j + \frac{1}{4}\sum_{ijkl}V_{ijkl}c^\dagger_i c^\dagger_j c_l c_k
\label{eq:H}
\end{equation}%
where $t$ and $V$ are the one-body energy and the two-body interaction
terms, respectively. $c^\dagger_i$ and $c_j$ are the
standard fermion creation and annihilation operators in a given orthonormal
single-particle basis $\{i,\ j,\ k,\ l,... \}$.

According to Eq.\ (\ref{eq:H}), the ground-state energy can be
expressed solely in terms of the second-order reduced density matrix,
2RDM, ${}^{2}\Gamma $ \cite{Husimi1940}
\begin{equation}
E_{0}[{}^{2}\Gamma ]=\frac{1}{4}\sum_{ijkl} H^{(2)}_{ijkl}\ ^2\Gamma_{ijkl}
\label{eq:Etr}
\end{equation}%
where
\begin{equation}
^2\Gamma_{ijkl} =\langle \psi |c_i^\dagger c_j^\dagger c_l c_k|\psi \rangle
\label{eq:Gamma}
\end{equation}
and
\begin{equation}
  H_{ijkl}^{(2)} =\frac{1}{N-1}\left( t_{ik}\delta_{jl} - t_{jk}\delta_{il}-t_{il}\delta_{jk}+t_{jl}\delta_{ik}\right) + V_{ijkl}
\end{equation}
is the two-particle reduced Hamiltonian with $|\psi \rangle $ the
ground-state wavefunction and $N$ the number of particles.

The idea behind the variational 2RDM methodology is to minimize the
energy functional (\ref{eq:Etr}) by varying the coefficients of
$^{2}\Gamma $. However, direct application of this procedure yields
unrealistic energies \cite{Coleman2000,Mayer1955,Lowdin1955} as
${}^{2}\Gamma $ must be constrained to the class of $N$-representable
2RDMs \cite{Coleman1963}. $N$-representability of a 2RDM implies there must exist an $N$-particle
wavefunction (or an ensemble of wavefunctions) from where it
derives. The necessary and sufficient conditions to assure the
$N$-representability of a $p$RDM are formally known
\cite{Garrod1964,Kummer1967,Coleman1974}: A $p$RDM is
$N$-representable if and only if for every $p$-body Hamiltonian
$H_{\xi }$ the following inequality is satisfied
\begin{equation}
\frac{1}{(p!)^2}\sum_{i_1 i_2 \cdots i_{2p}} H_{\xi\;i_1 i_2 \cdots i_{2p}}^{(p)}\;^{p}\Gamma_{i_1 i_2 \cdots i_{2p}} \geq E_{0}(H_{\xi })
\end{equation}%
with $H^{(p)}_\xi$ and $E_{0}(H_{\xi })$ being the $p$-particle
reduced Hamiltonian and the exact ground-state energy of $H_{\xi }$,
respectively. Unfortunately, this theorem cannot be used in practice
since it would require knowledge of the ground-state energy of every possible
$p$-body Hamiltonian $H_{\xi }$. However, it can be relaxed using a set of  Hamiltonians for
which a lower bound for the ground-state energy is known. This
is the case of the group of all semidefinite Hamiltonians, which are
completely defined by its extreme elements
\begin{equation}
H=B^{\dagger } B
\end{equation}%
yielding the well-known $\mathcal{P}$, $\mathcal{Q}$ and $\mathcal{G}$
two-index $N$-representability conditions
\cite{Coleman1963,Garrod1964} on the $^{2}\Gamma $ matrix if $%
B$ is restricted to the forms
$B=\sum_{ij} p_{ij} c_i c_j$,
$B=\sum_{ij} q_{ij} c^\dagger_i c^\dagger_j$, and
$B=\sum_{ij} g_{ij} c^\dagger_i c_j$, respectively. It has been shown that these conditions
are the necessary and sufficient conditions to assure the
$N$-representability for one-body
Hamiltonians \cite{Coleman1963,Coleman2000b}, as well as for two-body
Hamiltonians with an exact antisymmetric geminal power (AGP)\ ground
state \cite{Coleman2000b,Erdhal1974}. We will demonstrate this last
assertion analytically and numerically in Section \ref{sec:rgk_model} for the case of the
Richardson-Gaudin-Kitaev Hamiltonian.

Hamiltonians of the class
\begin{equation}
H=B^{\dagger }B+B B^{\dagger }
\end{equation}%
with
$B=\sum_{ijk}t_{ijk
}^{1}c_{i }^{\dagger }c_{j }^{\dagger }c_{k }^{\dagger
}$, and
$B=\sum_{ijk}t_{ijk}^{2}c_{i}^{\dagger }c_{j}^{\dagger }c_{k}$ yield the
$\mathcal{T}_{1}$ and $\mathcal{T}_{2}$ three-index
$N$-representability conditions
\cite{Erdhal1978,Zhao2004,Mazziotti2005} coming from the 3RDM on the
$^{2}\Gamma$ matrix, respectively. As these conditions are in
general necessary but not sufficient, the v2RDM will
always find a lower bound to the exact ground-state energy and an
approximation to the exact ground-state 2RDM.

In this work we will focus our attention on Hamiltonians with pairing
interactions in the seniority zero subspace. Assuming time-reversal
symmetry, the single-particle levels are doubly degenerate in the spin
degree of freedom. The seniority quantum number classifies the Hilbert space
into subspaces with a given number of singly occupied levels. The most
general pairing Hamiltonian conserving seniority is
\begin{equation}
H=\frac{1}{2}\sum_{i=1}^{L}\epsilon_i N_i +\sum_{ij=1}^{L}V_{ij}\,c^\dagger_i c^\dagger_{\bar{i}} c_{\bar{j}} c_j \label{Ham}
\end{equation}%
where $\epsilon_i$ are the energies of $L$ doubly degenerate
single-particle levels,
$N_i=c_i^{\dagger }c_i+c_{\bar{i}}^{\dagger}c_{\bar{i}}$ is the number
operator, and $V_{ij}$ is the pairing interaction. The $(i,\bar{i})$
pair defines the pairing scheme, which can involve two
particles with either opposite spins $(i\uparrow ,i\downarrow $), momenta
$(i,-i)$, or in general any classification of conjugate quantum
numbers in doubly degenerate single-particle levels. For these
Hamiltonians the seniority number is an exact quantum number, as
unpaired particles do not interact with the rest of the system and the
pairing Hamiltonian does not allow for pair breaking. The Hamiltonian
thus becomes block diagonal in sectors labeled by the seniority
quantum number.

The pairing Hamiltonian (\ref{Ham}) is based on the SU(2) pair algebra with
generators

\begin{equation}
K_i^+=c_i^\dagger c_{\bar{i}}^\dagger = \left(K_i^-\right)
^\dagger,\ K_i^z=\frac{1}{2}\left( N_i-1\right)
\end{equation}%
and commutation relations

\begin{equation}
\left[ K_i^+,K_j^-\right] =2\delta _{ij}K_i^z,\ \left[ K_i^z,K_j^\pm\right] =\pm \delta_{ij}K_i^\pm
\end{equation}%
We note that in the seniority zero subspace $N_i = 2K_i^+ K_i^-$ and
therefore, the Hamiltonian (\ref{Ham}) can be written in terms of the
ladder SU(2) operators as
\begin{equation}
H=\sum_{ij=1}^L J_{ij}K_{i}^{+}K_{j}^{-}
\end{equation}
where $J_{ij}=\delta _{ij}\epsilon _{i}+V_{ij}$. The ground-state energy is
thus given by
\begin{equation}
E_{0}=\sum_{ij=1}^L J_{ij}P_{ij}
\end{equation}%
where the $P$ matrix is
\begin{equation}
P_{ij}=\langle \psi |c_{i}^{\dagger }c_{\overline{i}}^{\dagger }c_{\overline{%
j}}c_{j}|\psi \rangle =\langle \psi |K_{i}^{+}K_{j}^{-}|\psi \rangle
\end{equation}%
This matrix together with the $D$ matrix
\begin{equation}
D_{ij}=\frac{1}{4}\langle \psi |N_{i}N_{j}|\psi \rangle =\langle \psi
|\left( K_{i}^{z}+\frac{1}{2}\right) \left( K_{j}^{z}+\frac{1}{2}\right)
|\psi \rangle
\end{equation}
define the seniority blocks of the $^{2}\Gamma$ matrix. Notice that the diagonal elements of both matrices are equal ($D_{ii} =P_{ii}$).  According to
these definitions, it follows that the $P$ and $D$ matrices are
hermitian and fulfill
\begin{align}
\sum_{i=1}^{L}P_{ii}& =\sum_{i=1}^LD_{ii}=M \label{eq:Mcond}\\
\sum_{j=1}^{L}D_{ij}& =MP_{ii} \label{eq:sumPD}
\end{align}%
where $M$ is the number of particle pairs in a system with $L$
doubly degenerate single-particle levels and total
$\langle \psi |K^{z}|\psi \rangle=M-\frac{L}{2}$. The $\mathcal{P}$,
$%
\mathcal{Q}$, $\mathcal{G}$, $\mathcal{T}_{1}$ and $\mathcal{T}_{2}$
$N$-representability conditions can thus be written in terms of the
seniority blocks of the 2RDM as
\cite{Weinhold1967,Weinhold1967b,Poelmans2015Thesis,Poelmans2015,Head-Marsden2017,Alcoba2018},

\begin{itemize}
\item The $\mathcal{P}$ condition:
\begin{align}
P& \succeq 0 \\
D_{ij}& \geq 0,\quad \forall i,j
\label{eq:Pcond}
\end{align}

\item The $\mathcal{Q}$ condition:
\begin{align}
Q& \succeq 0 \\
{\ }q_{ij}& \geq 0,\quad \forall i,j
\label{eq:Qcond}
\end{align}%
where
\begin{align}
Q_{ij}& =P_{ij}+\delta _{ij}(1-2P_{ii}) \\
q_{ij}& =D_{ij}+1-P_{ii}-P_{jj}
\end{align}

\item The $\mathcal{G}$ condition:
\begin{align}
G_{ij} &\succeq 0,\quad \forall i>j \\
g  & \succeq 0
\label{eq:Gcond}
\end{align}
where
\begin{align}
G_{ij} &=\left (
\begin{array}{cc}
P_{ii} - D_{ij} & -P_{ij} \\
-P_{ji} & P_{jj} - D_{ij}%
\end{array}%
\right) \\
g_{ij} &= D_{ij}
\end{align}

\item The $\mathcal{T}_1$ condition:
\begin{align}
T_1^i &\succeq 0, \quad \forall i \\
t_{1\,ijk} & \ge 0, \quad \forall i>j>k
\label{eq:T1cond}
\end{align}
where
\begin{align}
(T_1)^i_{jk} &= \delta_{jk}(1-2P_{jj}-P_{ii}+2 D_{ij}) + P_{jk},\quad
\forall j, k \neq i \\
t_{1\,ijk} &= 1 - P_{ii} - P_{jj} - P_{kk} + D_{ij} + D_{jk} + D_{ki}
\end{align}

\item The $\mathcal{T}_{2}$ condition:
\begin{align}
T_2^i& \succeq 0,\quad \forall i \\
T_{2}^{ijk}& \succeq 0,\quad \forall i>j>k
\end{align}%
where
\begin{equation}
T^i_2 =\left(
\begin{array}{ccc}
D_{jk} & -\delta _{jk}P_{ik} & D_{ik} \\
-\delta _{jk}P_{ki} & \delta _{jk}(P_{ii}-2D_{ik})+P_{jk} & P_{ik} \\
D_{ji} & P_{ji} & P_{ii}%
\end{array}%
\right)
\end{equation}
\begin{equation}
T_{2}^{ijk} =\left(
\begin{array}{ccc}
P_{ii}-D_{ij}-D_{ik}+D_{jk} & P_{ij} & P_{ik} \\
P_{ji} & P_{jj}-D_{ij}-D_{jk}+D_{ik} & P_{jk} \\
P_{ki} & P_{kj} & P_{kk}-D_{ik}-D_{jk}+D_{ij}
\label{eq:T2cond}
\end{array}%
\right)
\end{equation}

\end{itemize}
where the symbol  $\succeq 0$ denotes that a matrix is positive semidefinite.

The variational optimization of the 2RDM subject to conditions
(\ref{eq:Mcond})-(\ref{eq:T2cond}) can be formulated as a semidefinite
problem (SDP) in which the energy, being a linear function of the 2RDM, is minimized over the
intersection of a linear affine space and the convex cone of
block-diagonal positive semidefinite matrices
\cite{Nesterov1993,Vandenberghe1996,Wright1997,Wolkowicz2000}. As
discussed in \cite{Poelmans2015,Head-Marsden2017,Alcoba2018}, the SDP
in the seniority subspace computationally scales as $O(L^3)$ for the $\PQG$ conditions
and as $O(L^4)$ for the $\PQGT$ conditions. This will allow us to
treat without excessive computational efforts systems of sizes up to
$L=100$. In our numerical calculations we use the semidefinite programming algorithm (SDPA) code
\cite{sdpa-family,sdpa-v7}. This code
solves semidefinite problems at several precision levels by means of
the Mehrotra-type predictor-corrector primal-dual interior-point
method, providing ground-state energies and the corresponding 2RDM.

We programmed our v2RDM method as a  dual problem in the SDPA code, which does not allow for the equality constraints (\ref{eq:Mcond})-(\ref{eq:sumPD}). These are included by relaxing them into inequality constraints with a sufficiently small summation error\ $\delta$\ \cite{Zhao2004, Nakata2008}. In our work we have set\ $\delta = 10^{-7}$, which effectively fixes the precision of the ground-state energies.

\section{Richardson-Gaudin integrable models}

The Richardson-Gaudin (RG) models are based on a set of integrals of motion (IM)\ or
quantum invariants that are linear and quadratic combinations of the
generators of the\ SU(2) algebra. By requiring the IM to commute with
the total spin operators $K^{z}=\sum_{i}K_{i}^{z}$, the most general
expression for the IM is

\begin{equation}
R_{i}=K_{i}^{z}+G\sum_{j\left( \neq i\right) }\frac{X_{ij}}{2}\left(
K_{i}^{+}K_{j}^{-}+K_{j}^{+}K_{i}^{-}\right) +Z_{ij}K_{i}^{z}K_{j}^{z}
\label{eq:IM}
\end{equation}%
where $X_{ij}$ and $Z_{ij}$ are antisymmetric matrices and $G$ is the
pairing strength. The operators $R_{i}$ must commute among themselves
to constitute a set of IM. Imposing these conditions leads to two
families of integrable models:

\begin{enumerate}
	\item The hyperbolic or XXZ family
	
	\begin{equation}
	X_{ij}=\frac{2\eta_i\eta_j}{\eta^2_i-\eta^2_j},\
	Z_{ij}=\frac{\eta^2_i+\eta^2_j}{\eta^2_i-\eta^2_j}
	\end{equation}%

	\item The rational or XXX family
	
	\begin{equation}
	X_{ij}=Z_{ij}=\frac{1}{\eta^2_i-\eta^2_j}
	\end{equation}%
		
\end{enumerate}
where the $\eta ^{\prime }s$ are an arbitrary set of real parameters.

The common eigenstates of IM (\ref{eq:IM}) are determined by the solution of the set of $M$ non-linear coupled RG equations
\begin{equation}
	1 + \frac{G}{2} \sum_{i=1}^{L} Z_{i\alpha} - G\sum_{\beta\neq\alpha}^{M}Z_{\beta\alpha} = 0,\quad \forall\ \alpha=1,\cdots,M
\label{eq:RG}
\end{equation}
with\ $Z_{i\alpha} = Z\left(\eta^2_i, E_\alpha \right)$ in terms of the $M$ spectral parameters \ $E_\alpha$.

Defining the new $L$ variables
\begin{equation}
\Lambda_i = \sum_{\alpha=1}^{M} Z\left(\eta^2_i, E_\alpha \right)
\end{equation}
we can write the RG equations as a set of $L$ coupled quadratic equations \cite{Pieter2015} in the $\Lambda$ variables
\begin{equation}
	\Lambda^2_i = M(L-M)C - \frac{2}{G} \Lambda_i + \sum_{j\neq i}^{L} Z_{ij}\left(\Lambda_i - \Lambda_j\right)
\end{equation}
where\ $C$ is a constant that depends on the Gaudin algebra,\ $0$ for the rational family and\ $-1$ for the hyperbolic family. This new system of equations is free of the divergences that plague the original set of RG equations (\ref{eq:RG}), and it can be solved numerically with the Levenberg-Marquardt algorithm. Once we have determined the set of\ $\Lambda_i$ for a particular eigenstate, the eigenvalues of the IM are
\begin{equation}
	r_i = \frac{1}{2}\left(-1 -G\Lambda_i + \frac{G}{2}\sum_{j\neq i}^{L}Z_{ij} \right)
\end{equation}
If the Hamiltonian is an arbitrary linear combination of the IM, \ $H = \sum_{i=1}^L \varepsilon_i R_i$, the corresponding eigenvalue is
\begin{equation}
	E = \sum_{i=1}^L \varepsilon_i r_i
\end{equation}

\subsection{The Richardson-Gaudin-Kitaev model}
\label{sec:rgk_model}

The Richardson-Gaudin-Kitaev (RGK) model \cite{Ortiz} is a variation of the
celebrated Kitaev wire \cite{Kitaev} proposed as a toy model to understand
topological superconductivity. While the Kitaev wire is a
non-number-conserving one-body Hamiltonian for spinless fermions in a
1D chain, the RGK Hamiltonian is two-body and number
conserving. Moreover, it is exactly solvable for closed boundary
conditions, either periodic or antiperiodic. Hence, this interacting
  many-body Hamiltonian allows to obtain precise answers for the
  characterization of topological superconductivity.

The RGK\ Hamiltonian is a particular realization of the hyperbolic family of
RG models describing p-wave pairing \cite{Ibanes, Rombouts, Neck}
\begin{equation}
	H=\frac{1}{2}\sum_{i\in I+} \varepsilon_i N_i - G\sum_{ij \in I+} \eta_i\eta
	_j c^\dagger_i c^\dagger_{\bar{i}} c_{\bar{j}} c_{j}
	\label{eq:rgk_hamiltonian}
\end{equation}
where $\eta_i=\sin \left( i/2\right)$ and
$\varepsilon_i=\eta_i^{2}$, such that the one-body term
describes near-neighbor hoppings in a 1D chain. For simplicity we will
assume antiperiodic boundary conditions.  In this case, the allowed
values of $i$ in a 1D chain of length $2L$ are
$I+=\left\{ \pi , 3\pi ,\cdots ,\left( 2\pi L-\pi
  \right) \right\} /(2L)$.

The complete set of eigenstates in the seniority zero subspace is given by a
product pair ansatz
\begin{equation}
\left\vert \Psi \right\rangle =\prod\limits_{\alpha =1}^{M}\left(
\sum_{i=1}^{L}\frac{\eta_i}{\varepsilon_i-E_{\alpha }}c^\dagger_i c^\dagger_{\bar{i}}\right)
\left\vert 0\right\rangle
\label{PsiRGK}
\end{equation}%
where the set of $M$ spectral parameters (pair energies) $E_{\alpha }$ are a
particular solution of a set of $M$ non-linear coupled RG equations
and $|0\rangle$ is the vacuum  state.

The ground state solution has two critical values of $G$ with peculiar
properties: the Moore-Read point $G_{MR}=\frac{1}{L-M+1}$ \cite{Moore-Read}, and the
Read-Green point $G_{RG}=\frac{1}{L-2M+2}$ \cite{Read-Green}.

For the ground state solution at the Moore-Read point $G_{MR}$, and
independently of the definition of the $\eta ^{\prime }s$, all pair energies
$E_{\alpha }$ collapse at 0 energy
($E_{\alpha }=0,\ \forall\ \alpha $). Therefore, the RGK\ ground state
for $G_{MR}$ is a pair condensate also known as number projected BCS
(PBCS) wavefunction in nuclear physics or antisymmetric geminal power,
AGP, in quantum chemistry
\begin{equation}
\left\vert \Psi _{MR}\right\rangle =\left( \sum_{i=1}^{L}\frac{1}{\eta_i}%
c^\dagger_i c^\dagger_{\bar{i}}\right)^M\left\vert 0\right\rangle
\label{PsiMR}
\end{equation}

The PBCS or AGP\ wavefunction, being exact at $G_{MR}$, will display
important consequences for the v2RDM approach. As mentioned above, the
$\PQG$ conditions are sufficient to produce the exact v2RDM\ result at
this point. This statement can be independently proven starting from
the set of killers of an\ AGP wavefunction
\begin{equation}
B_{ij}=\eta _{j}c_{i}^{\dagger }c_{j}-\eta _{i}c_{\overline{j}}^{\dagger }c_{%
\overline{i}}
\end{equation}%
such that
\begin{equation}
B_{ij}\left\vert \Psi _{MR}\right\rangle =0  \label{Kill}
\end{equation}
from which the Moore-Read Hamiltonian derives as the positive semidefinite
operator
\begin{eqnarray}
	H_{MR} &=&\frac{1}{\left[ 2\left( L+1\right) -2M\right] }\sum_{ij}B_{ij}^{%
	\dagger }B_{ij} \\
	&=& \frac{1}{2} \sum_{i}\eta^2_i N_{i}-\frac{1}{L-M+1}\sum_{ij}\eta
	_{i}\eta _{j}c_{i}^{\dagger }c_{\overline{i}}^{\dagger }c_{\overline{j}}c_{j}
\label{HMR}
\end{eqnarray}%
with $0$ ground-state energy.

\begin{figure}[htbp]
\includegraphics[width=0.6\textwidth]{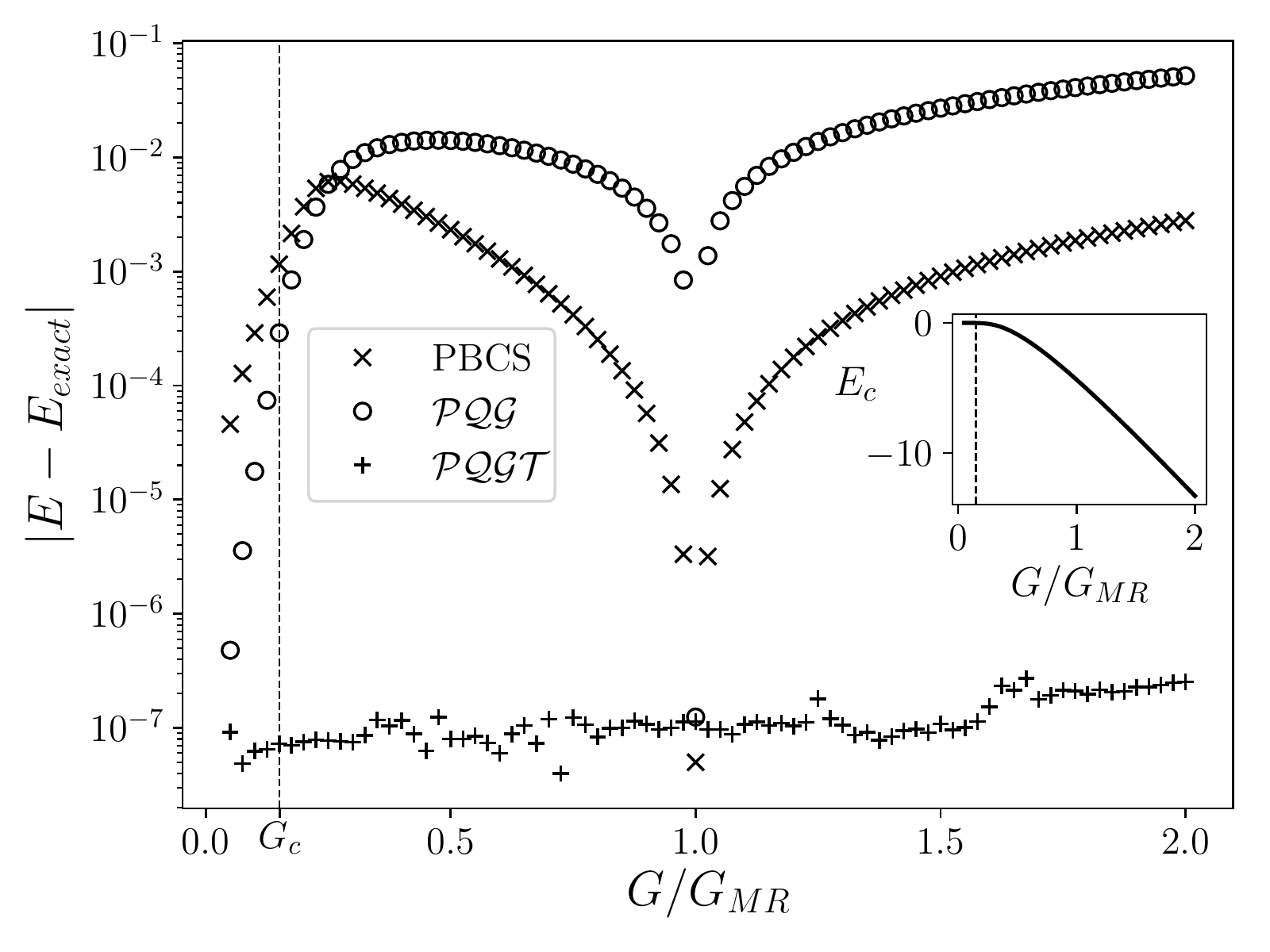}
\caption{Absolute energy difference of v2RDM and PBCS  with respect to the exact
  ground-state energy of the RGK Hamiltonian at
  different interaction strengths for a system with\ $L=50$ doubly degenerate
  levels. The v2RDM results are computed
  with the\ $\mathcal{PQG}$ and the\ $\mathcal{PQGT}$
  conditions. Inset: exact correlation energy.}
\label{fig:rgk_energy_diff}
\end{figure}

The Read-Green point $G_{RG\text{ }}$ signals the topological quantum
phase transition. In the thermodynamic limit the scaled pairing
strength is $%
g_{RG}=G_{RG} L=1/\left( 1-2\rho \right) $ implying that there is no
phase transition for densities $\rho \geq 1/2$. Since we are
interested in testing the accuracy of the v2RDM specifically around
the Moore-Read point, we will consider systems of different sizes at
half filling for several values of the pairing strength in units of
$G_{MR}$.

In addition to the ground-state energies we will test another magnitude that
characterizes the pair mixing across the Fermi level, the canonical gap
defined as
\begin{align}
\Delta_c&=G\sum_{i=1}^L \eta_i\sqrt{\left\langle c_i^\dagger c_{\bar{i}
}^\dagger c_{\bar{i}}c_i\right\rangle -\left\langle c_i^\dagger c_i\right\rangle \left\langle c_{\bar{i}}^\dagger c_{\bar{i}}\right\rangle}
\nonumber \\
&= G\sum_{i=1}^L \eta _i\sqrt{P_{ii}\left( 1-P_{ii}\right) }
\label{Gap}
\end{align}
It turns out that $\Delta _{c}$ coincides with the BCS gap $\Delta$
when it is evaluated with a number non-conserving BCS wavefunction. In
this case the BCS gap equation reduces to
\begin{equation}
\frac{1}{G}=\sum_{i=1}^L \frac{\eta_i}{\sqrt{\left( \varepsilon_i-\mu
\right)^{2}+\eta^2_i\Delta ^{2}}},\ \mu =\frac{\varepsilon
_{M}+\varepsilon _{M+1}}{2}
\end{equation}%
As a function of $G$ the system has a phase transition from a metallic
state characterized by $\Delta =0$ to a superconducting state with
finite gap. The critical value of $G$ is obtained from the gap
equations as

\begin{equation}
G_{c}=\left[ \sum_i\frac{\eta_i}{\left\vert \varepsilon_i-\mu
\right\vert }\right]^{-1}
\end{equation}

Even though BCS predicts a non-superconducting state for $G<G_{c}$ ($\Delta
=0$), for correlated number conserving wavefunctions like PBCS or AGP the
gap is always greater than zero \cite{Pittel}.

\begin{figure}[htbp]
\includegraphics[width=0.6\textwidth]{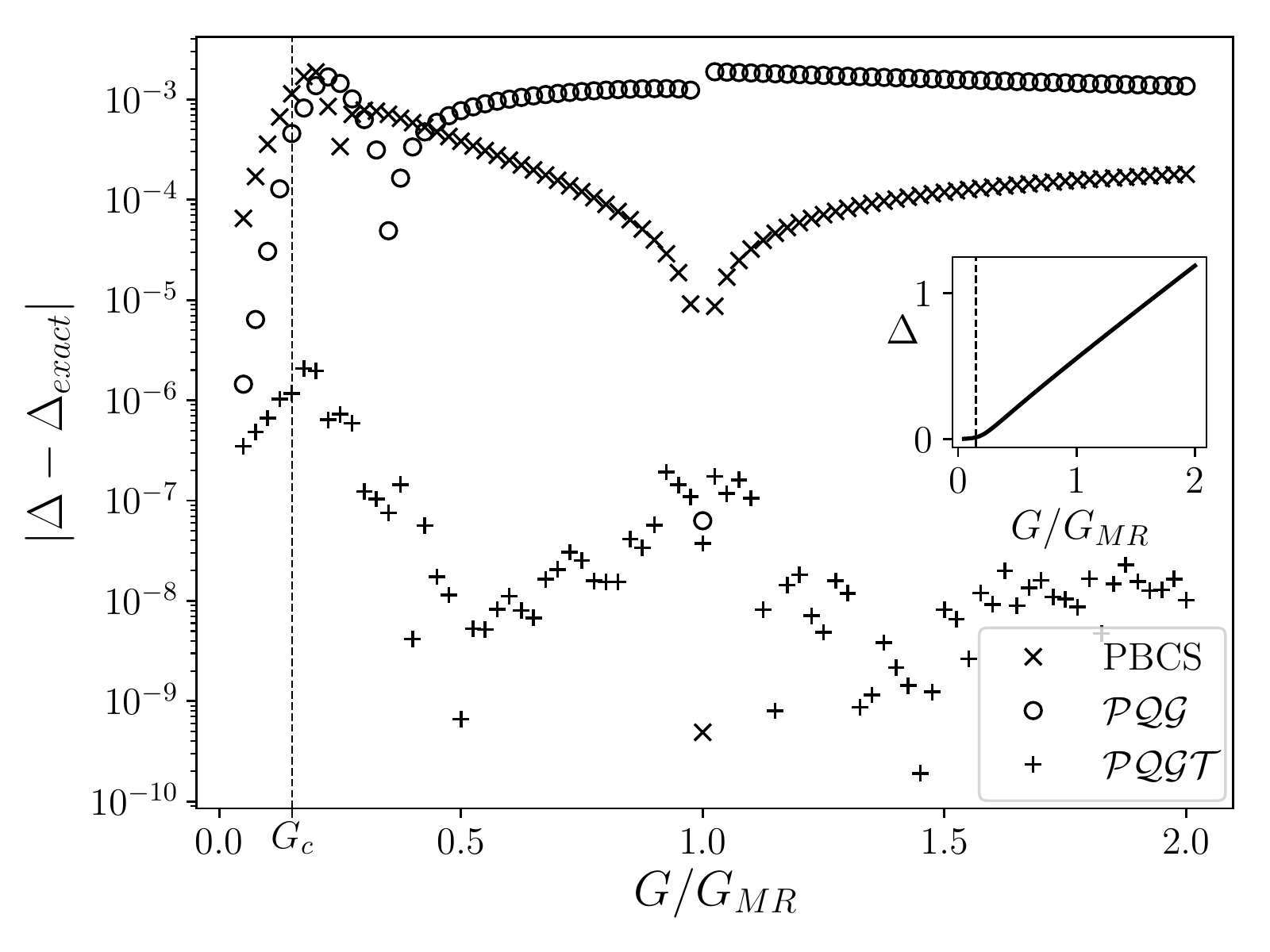}
\caption{Absolute canonical gap difference of the v2RDM and the PBCS with respect to the exact
  canonical gap of the RGK Hamiltonian at
  different interaction strengths. The v2RDM results are computed with the\ $\mathcal{PQG}$ and the\
  $\mathcal{PQGT}$ conditions. Computations are for a system with\
  $L=50$ doubly degenerate levels. Inset: exact canonical gap.}
\label{fig:rgk_gap_diff}
\end{figure}

We have now all the tools for testing the different variational
approximations with the exact solution of the RGK model. We start with
a system of $L=50$ doubly degenerate levels at half filling
corresponding to $M=25$ fermion pairs.  The size of the Hilbert space
is $1.26 \times 10^{14}$, well beyond the limits of an exact
diagonalization.  Note that for finite size systems at half-filling, the Read-Green point  lies at very large values of $G$,
$G_{RG}=1/2$ as compared to the Moore-Read point ($G_{MR}=1/26$).
 Therefore, we assume $G_{MR}$ as a characteristic value of the pairing strength, at which
PBCS and the\ $\mathcal{PQG}$ v2RDM approximations must be
exact. Thus, we will study the behavior of the different
approximations as a function of $G$ in units of $G_{MR}$.

Fig.\ \ref{fig:rgk_energy_diff} shows the absolute value of the
difference between the approximated and the exact ground-state energy. We
display here the absolute value in order to compare PBCS and
v2RDM. However, we should keep mind that this difference is positive
for PBCS due to its Ritz variational character, while it is negative
for v2RDM because it provides lower bounds. The inset displays the
behavior of the correlation energy, which stays flat for weak pairing, and starts to decrease linearly with $G$ entering the
superconducting region. The correlation energy is defined as
\begin{equation}
	E_c = \langle\psi|H|\psi\rangle - \langle\psi(0)|H|\psi(0)\rangle
\end{equation}
where\ $|\psi(0)\rangle$ is the ground state of the noninteracting Hamiltonian.

\begin{figure}[htbp]
\includegraphics[width=0.6\textwidth]{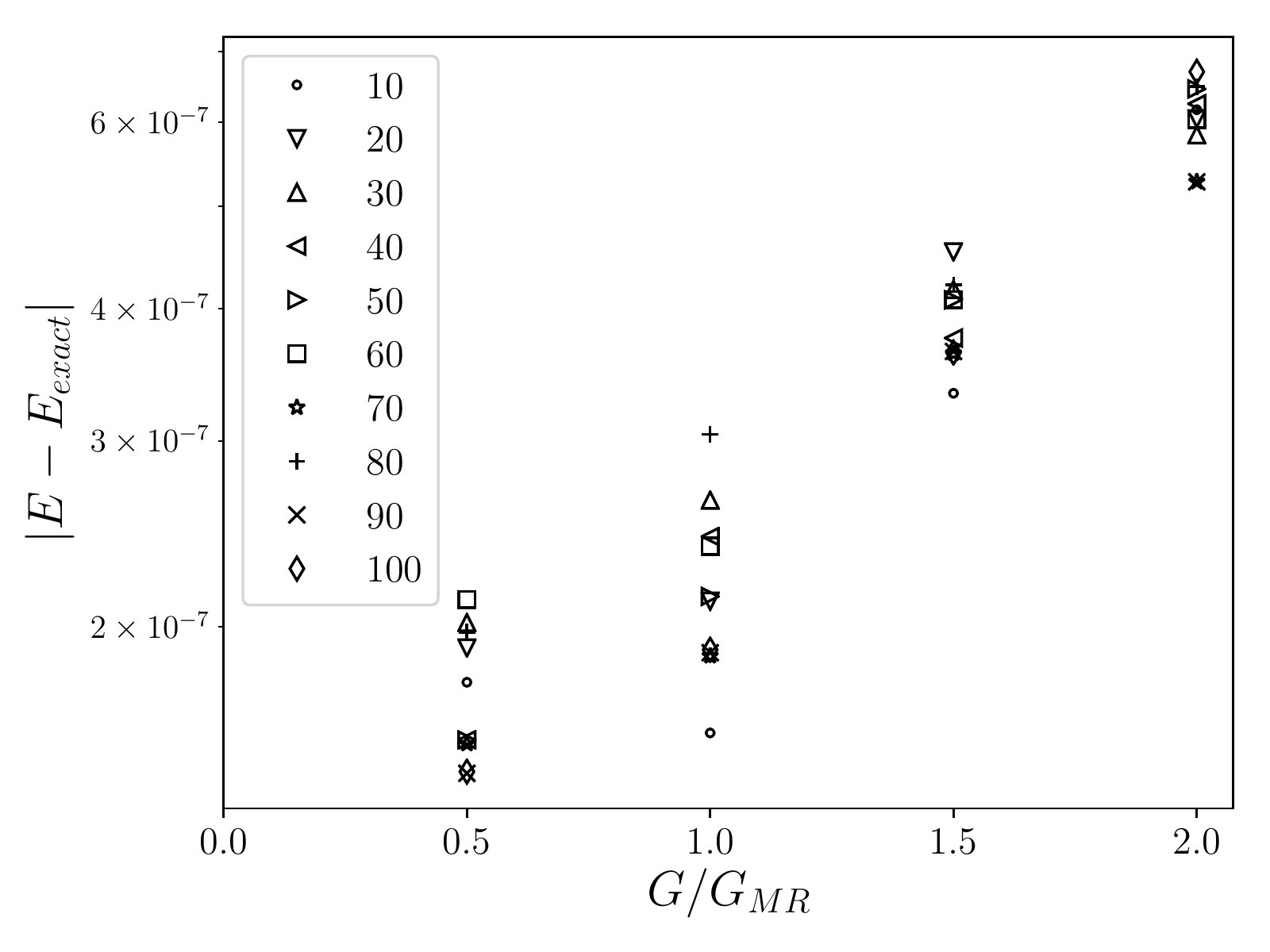}
\caption{Absolute energy difference of the v2RDM with the\
  $\mathcal{PQGT}$ conditions with respect to the exact ground-state
  energy of the RGK Hamiltonian with different number of doubly
  degenerate levels as specified in the legend)}
\label{fig:rgk_energy_various_levels}
\end{figure}

As it was expected, the\ $\mathcal{PQG}$ and PBCS are indeed exact at
the Moore-Read point \ $G_{MR}$. While both approximations have a
comparable accuracy in the weak coupling region, PBCS is two orders of
magnitude better in the superconducting region. In contrast, the
addition of constraints coming from the 3RDM in the\ $\mathcal{PQGT}$
approximation makes the formalism numerically exact within the
accuracy limit imposed by the semidefinite programming code SDPA.

Fig.\ \ref{fig:rgk_gap_diff} shows the comparison of the canonical
gap\ (\ref{Gap}) computed with PBCS and the v2RDM with the exact
one. As the gap is not determined from a variational principle, we
plot the absolute value of the differences between approximated and
exact gaps. Again, the PBCS and $\mathcal{PQG}$ gaps are exact at the
Moore-Read point, providing a second numerical confirmation of the
exactness of both approaches. The $\mathcal{PQG}$ approximation
manages to give a fairly good description of the gap but the PBCS
again provides at least one order of magnitude approximation better in
the superconducting region. The computations with the\
$\mathcal{PQGT}$ conditions give again a numerically exact
approximation to the canonical gap. The inset in the figure shows the
behavior of the exact canonical gap, which remains small at weak
interactions due to pairing fluctuations, until it opens at around the
critical interaction strength\ $G_c$, where the system enters a
superconducting phase.  The gaps in the PBCS and $\mathcal{PQG}$
approximations show some structure for
$G/G_c \sim 1.7$ and $2.3$ for which we could not
find an explanation. However, this structure disappears with the
$\PQGT$ conditions.

To ensure that the v2RDM method is extensible to systems of arbitrary
sizes we show in Fig.\ \ref{fig:rgk_energy_various_levels} the
comparison of the total ground-state energy under the\
$\mathcal{PQGT}$ conditions with the exact energy for systems with
sizes ranging from\ $L=10$ to\ $100$ levels. To compute systems of
such larger sizes we have relaxed the summation error to\ $\delta=3\cdot 10^{-7}$, which is marginally lower that the previous computations. Our results show that the
exact ground-state energies are numerically exact to the required
precision independently of the system sizes. The relative energy errors
are of the same order of magnitude taking into account that the
correlation energy (inset of Fig. \ref{fig:rgk_energy_diff}) increases
by one order of magnitude along the horizontal axis.

\subsection{The reduced BCS Hamiltonian}

The reduced BCS or constant pairing Hamiltonian has been widely
employed in condensed matter and nuclear physics to study
superconducting properties of extensive as well as finite size systems
in the BCS approximation. Few years after the celebrated BCS paper,
Richardson solved this Hamiltonian exactly \cite{Rich63}. More recently, the exact
solution has been generalized to families of exactly solvable pairing
models \cite{Dukelsky2001}. In this subsection we will resort to the constant pairing
Hamiltonian in the form used to describe ultrasmall superconducting
grains \cite{Delft96}
\begin{equation}
H_{BCS}=\sum_{i=1}^L \frac{\varepsilon _{i}}{2}N_{i}-G\sum_{ij=1}^L c_{i}^{\dagger }c_{%
\overline{i}}^{\dagger }c_{\overline{j}}c_{j}  \label{Hbcs}
\end{equation}
Richardson proposed a product pair ansatz for the exact eigenstates of the BCS
Hamiltonian
\begin{equation}
\left\vert \Psi \right\rangle =\prod\limits_{\alpha =1}^{M}\left(
\sum_{i=1}^{L}\frac{c_{i}^{\dagger }c_{\overline{i}}^{\dagger }}{\varepsilon
_{i}-E_{\alpha }}\right) \left\vert 0\right\rangle  \label{PsiBCS}
\end{equation}
As in the RGK\ case, the pair energies,\ $E_\alpha$, are
obtained from the solution of a set of $M$ nonlinear coupled equations and
the total eigenvalues are the sum of the pair energies $E=\sum_{\alpha=1}^M E_{\alpha }$.

\begin{figure}[htbp]
	\includegraphics[width=0.6\textwidth]{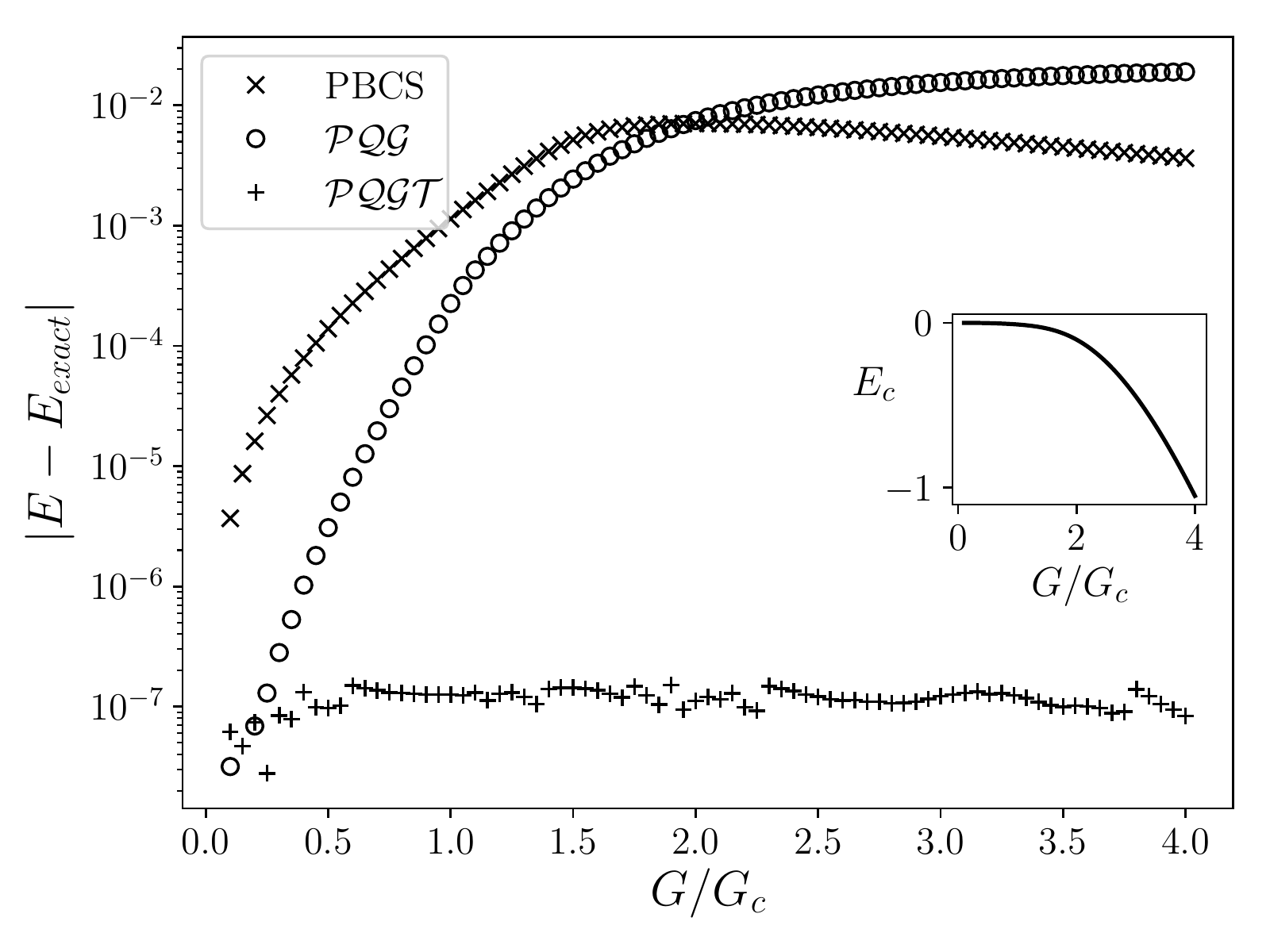}
	\caption{Absolute energy difference of the v2RDM and the PBCS
          with respect to the exact ground-state energy of the
          constant pairing Hamiltonian at different interaction
          strengths for a system with\ $L=50$ doubly degenerate
          levels. The v2RDM results are computed with the\
          $\mathcal{PQG}$ and the\ $\mathcal{PQGT}$ conditions.
          Inset: exact correlation energy.}
	\label{fig:pairing_energy_diff}
\end{figure}

Note the slight difference with the eigenstates of the RGK\
Hamiltonian. In spite of the similarities in the wavefunction, the
physics of these two Hamiltonians is completely different. While the
BCS Hamiltonian describes fermions interacting through an attractive
s-wave pairing, the RGK\ Hamiltonian describes a p-wave
interaction. In the former case there is a smooth crossover from a
superconducting BCS state to a Bose-Einstein condensate \cite{crossover}. In the latter
case there is a third-order quantum phase transition from a
topological superconducting phase to a trivial superconducting phase
or Bose-Einstein condensate of p-wave molecules \cite{Rombouts}.

\begin{figure}[htbp]
	\includegraphics[width=0.6\textwidth]{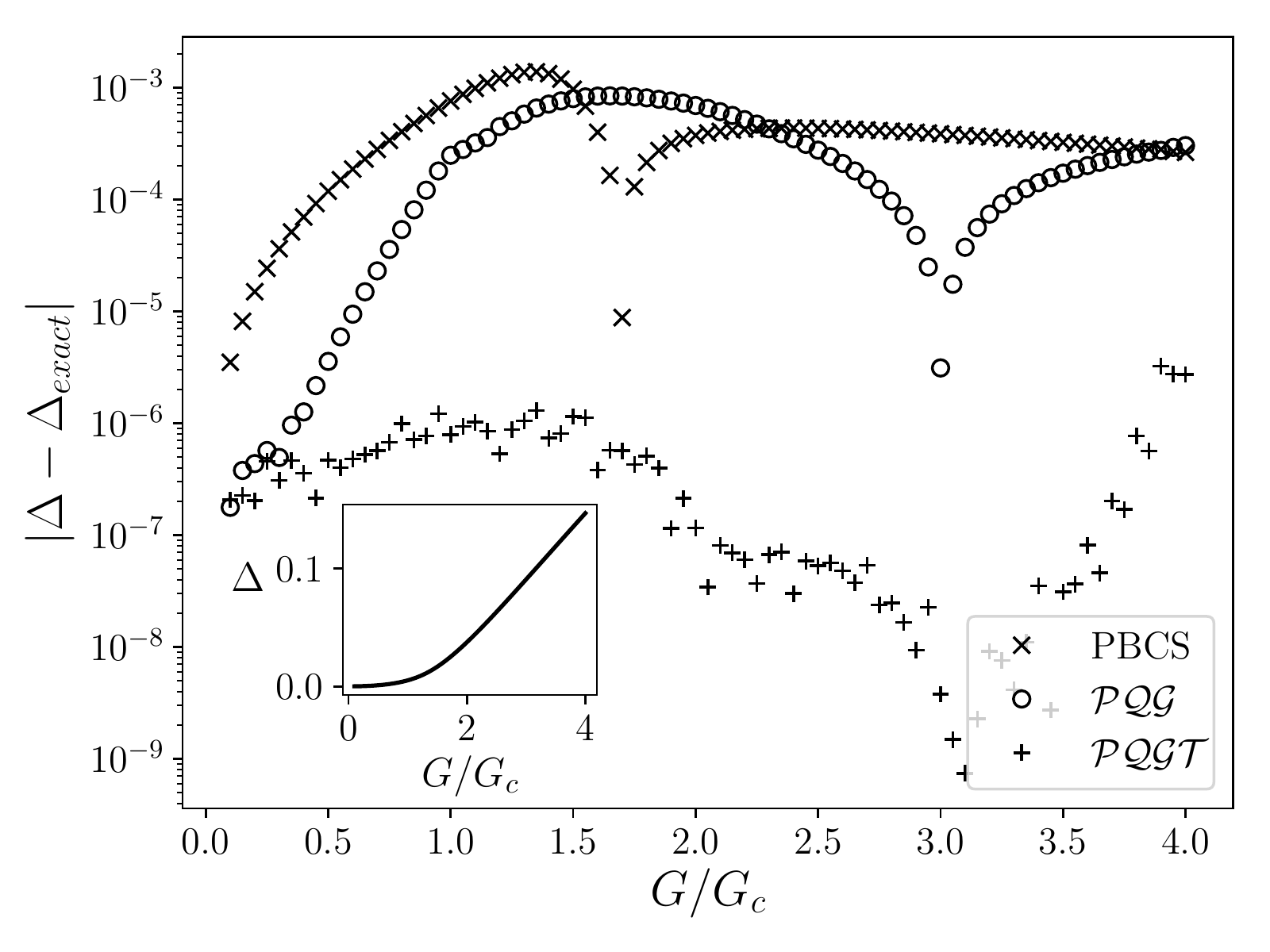}
	\caption{Absolute canonical gap difference of the v2RDM and
          the PBCS with respect to the exact canonical gap of the
          constant pairing Hamiltonian at different interaction
          strengths. The v2RDM results are computed with the\
          $\mathcal{PQG}$ and the\ $\mathcal{PQGT}$
          conditions. Computations are for a system with 50 doubly
          degenerate levels. }
	\label{fig:pairing_gap_diff}
\end{figure}

In small grains it is customary to assume equidistant levels and to
express all quantities in units of the mean level spacing $d$, which
in turn is inversely proportional to the volume of the grain. However,
due to presence of disorder, the level spacing in small metallic grains
follows a Wigner-Dyson distribution obtained from random matrix
theory. We will take advantage of the two standard descriptions of
small grains to benchmark the v2RDM. First, we will test it with
uniformly distributed equidistant levels, and then investigate how
robust is the method in the presence of random disorder.

In order to quantify pairing fluctuations around the Fermi level we make use
of the canonical gap $\Delta _{c}$
\begin{align}
\Delta_{c}&=G\sum_{i=1}^L \sqrt{\left\langle c_{i}^{\dagger }c_{\overline{i}%
}^{\dagger }c_{\overline{i}}c_{i}\right\rangle -\left\langle c_{i}^{\dagger
}c_{i}\right\rangle \left\langle c_{\overline{i}}^{\dagger }c_{\overline{i}%
}\right\rangle } \nonumber \\
&=G\sum_{i=1}^L \sqrt{P_{ii}\left( 1-P_{ii}\right) }
\end{align}
For finite systems the BCS approximation has a metallic phase with no
gap, and a superconducting phase with finite gap. The critical value
of\ $G$ is
\begin{equation}
G_{c}=\left[ \sum_{i}\frac{1}{\left\vert \varepsilon _{i}-\mu \right\vert }%
\right] ^{-1}
\end{equation}

Since $G_{c}$ is a sensible value to assess the degree of
superconducting correlations, we will study the BCS Hamiltonian for
different system sizes as a function of $G$ \ in units of $G_{c}$.

\begin{figure}[htbp]
	\includegraphics[width=0.6\textwidth]{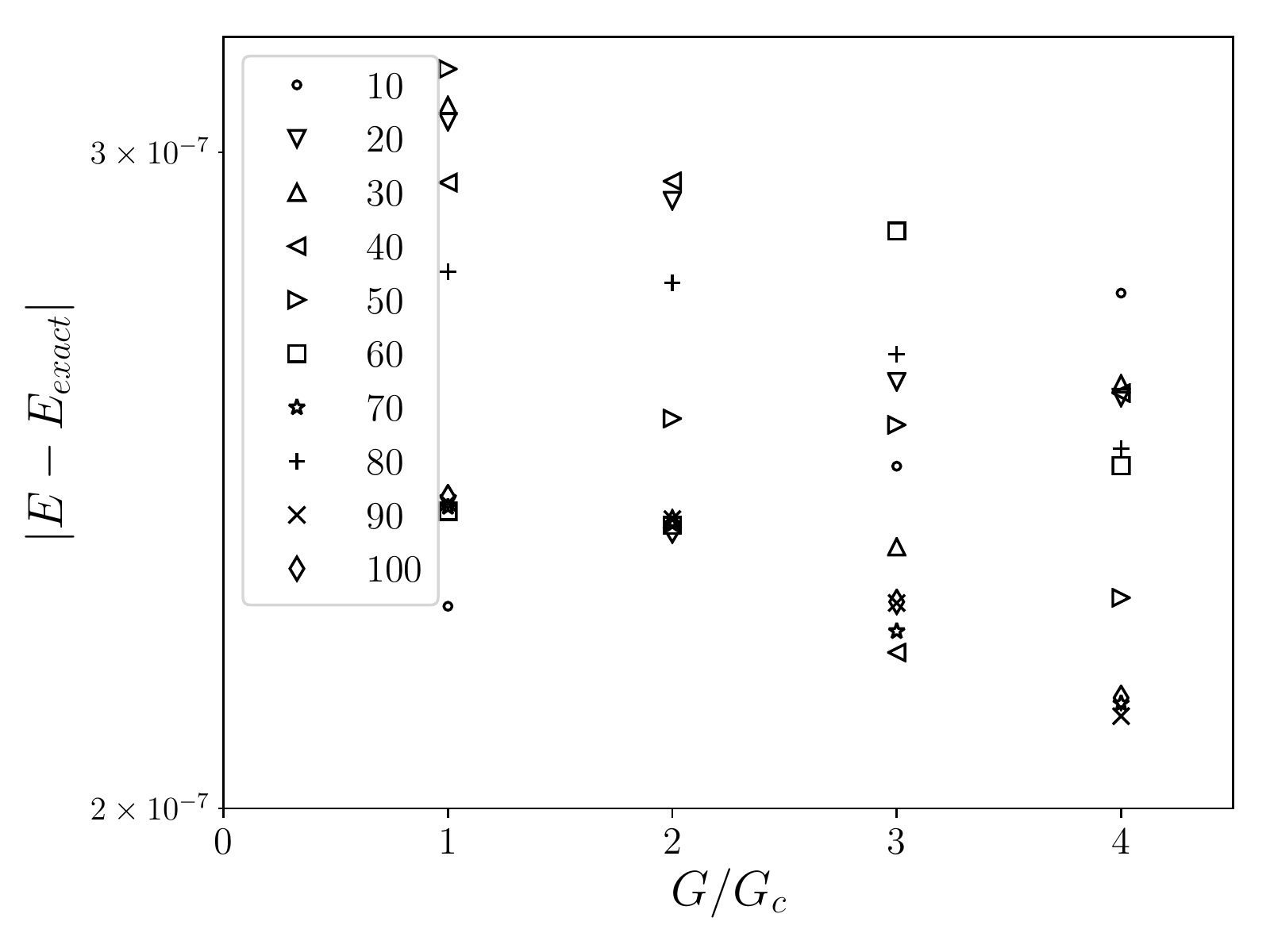}
	\caption{Absolute energy difference of the v2RDM with the\
          $\mathcal{PQGT}$ conditions with respect to the exact
          ground-state energy of the constant pairing Hamiltonian with
          different number of doubly degenerate levels as specified in
          the legend.}
	\label{fig:pairing_energy_various_levels}
\end{figure}

Fig.\ \ref{fig:pairing_energy_diff} shows the absolute value of the
differences between the ground-state energy in the different
approximations and the exact one for a system of $M=25$ fermion pairs
in $L=50$ equidistant single-particle levels with spacing $d=1/L$
as a function of the pairing strength $G$ in units of $G_c$. As in the
case of the RGK model, the $\mathcal{PQGT}$ conditions are sufficient
enough to reproduce the exact results within the numerical error of
the computing program. Ref. \cite{Exact} found the same conclusion for
a system of $M=12$ pairs.  $\mathcal{PQG}$ and PBCS are significantly
less precise with a complementary behavior. $\mathcal{PQG}$ starts
with a good description of the system at weak pairing, but it quickly
degrades approaching the critical region. On the contrary, PBCS is
less accurate in weak pairing but tends to improve towards the strong
superconducting region.  The inset displays the exact correlation
energy as a function of $G$, exhibiting a change in curvature around
the critical BCS value of $G$ that separates a regime dominated by
pairing fluctuations from a superconducting phase characterized by a
condensation of Cooper pairs. A similar picture is described in the
inset of Fig.\ \ref{fig:pairing_gap_diff} with small but nonzero
values of the canonical gap below $G_c$ changing to a linear behavior
above $G_c$.

Fig.\ \ref{fig:pairing_gap_diff} confirms the remarkable accuracy of
the $\mathcal{PQGT}$ approximation. Curiously, the gaps in the PBCS
and $\mathcal{PQG}$ approximations show a similar behavior for
$G/G_c \sim 1.7$ and $3.0$ respectively as in the RGK model.

Fig.\ \ref{fig:pairing_energy_various_levels} explores the accuracy of
the $\PQGT$ method as a function of the system size in a similar way
as it has been done for the RGK Hamiltonian. As seen in the figure,
the v2RDM energies are numerically exact within the accepted
tolerance. The relative errors are comparable since the correlation
energy is of the same order of magnitude for the whole range of
interactions (inset of Fig. \ref{fig:pairing_energy_diff}). As in the RGK example, we have relaxed the summation error to\ $\delta=3\cdot 10^{-7}$.

\begin{figure}[htbp]
	\includegraphics[width=0.6\textwidth]{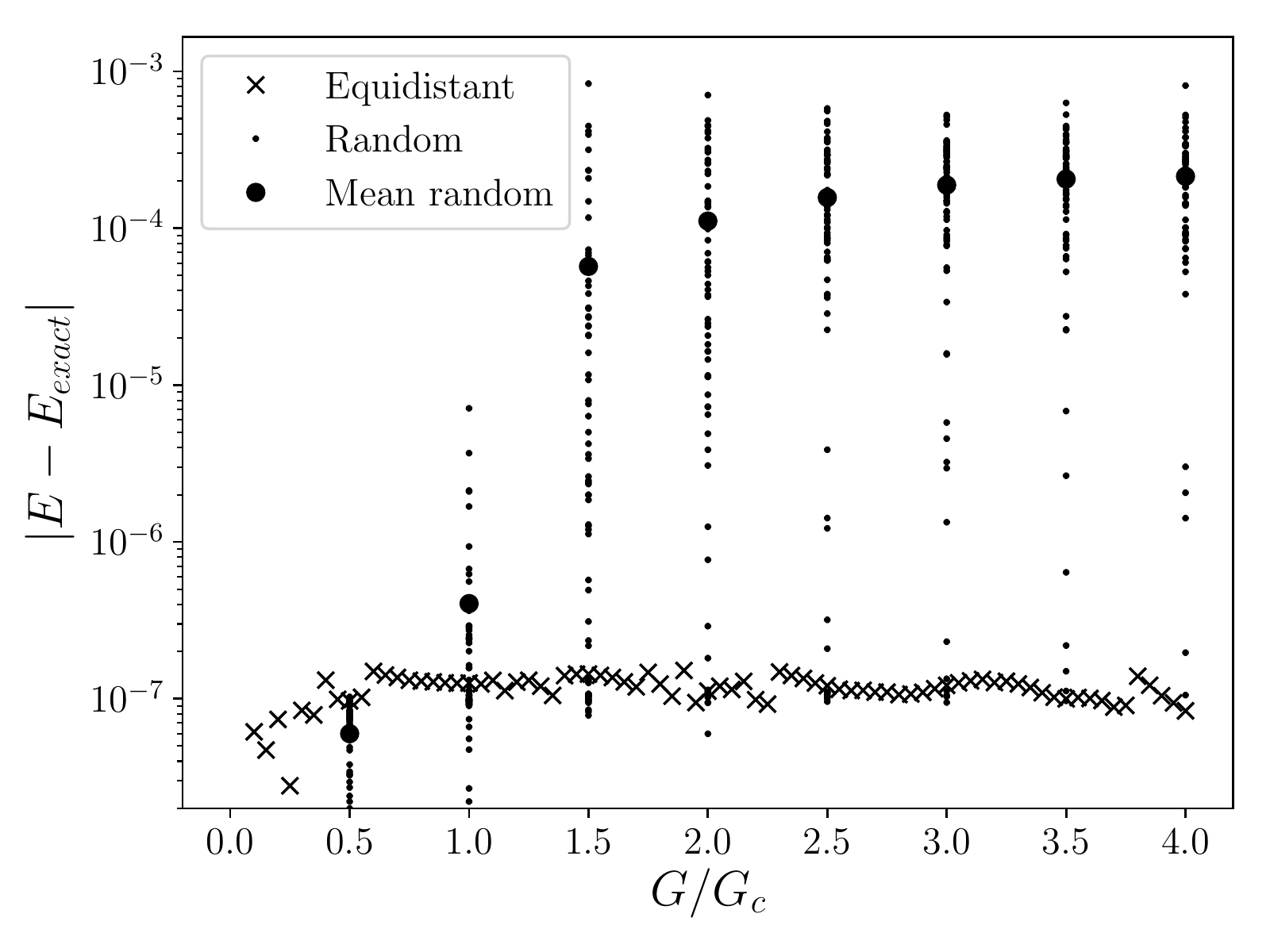}
	\caption{Absolute energy differences of the v2RDM with the
          $\PQGT$ conditions of 70 samples of random single-particle
          energies with respect to the exact ground-state for selected
          values of $G/G_c$. The big solid circle signals the mean
          value of the ensemble. For comparison we display the $\PQGT$
          energies of the equidistant single-particle
          case. Computations are for a system with 50 doubly
          degenerate levels.}
	\label{fig:random_pairing_energy_diff}
\end{figure}

It is known that the energy levels of small metallic grains follow a
Gaussian orthogonal ensemble distribution. For simplicity, most of the
studies have been carried out assuming a uniform level
spacing. However, the exact solution of the BCS Hamiltonian
(\ref{Hbcs}) is valid for arbitrary single-particle levels
$\varepsilon_i$. This feature has been exploited to study in an exact manner the
interplay between randomness and
interaction in the crossover from metal to superconductor as a
function of the grain size \cite{Delft}. Here, we will use this
ability of the exact solution to test the robustness of the
$\mathcal{PQGT}$ conditions against disorder in the single-particle levels
spectrum. For each value of $G/G_c$ in Fig.\
\ref{fig:random_pairing_energy_diff} we generate 70 symmetric random
matrices of size $2L\times 2L$. Upon diagonalization, we select the
central $L$ eigenvalues to avoid edge effects. In order to assure an
average constant level spacing we rescale them as
$ \varepsilon \rightarrow  (1 / 2\pi) [4 L \sin^{-1} \left(\varepsilon / \sqrt{4 L}\right) +
\varepsilon \sqrt{ 4 L - \varepsilon^2}  ] $.

Fig.\ \ref{fig:random_pairing_energy_diff} shows the results obtained
in the $\mathcal{PQGT}$ approximation for each random ensemble as
compared with the uniform level spacing case. Interestingly enough, the
transition from metallic to superconductor reveals a clear cut
distinction in the accuracy of the v2RDM method. While the method is
completely accurate for all instances below the critical $G_c$ value,
it starts to deviate from the exact ground-state energy crossing this
point and loosing three orders of magnitude in accuracy.  In spite of
this loss, errors of $10^{-4}$ in the correlation energy are quite
acceptable for many standards. However, the reason of this deviation
cannot be attributed the loss of integrability since the random
Hamiltonian (\ref{Hbcs}) is always exactly solvable and the exact
eigenstates are given by the ansatz (\ref{PsiBCS}). It might be
attributed to the complexity of the wavefunction (\ref{PsiBCS}) with
random energy levels $\varepsilon_i$.

\section{Summary}

In this work we have explored the performance of the v2RDM within the
seniority zero subspace for two classes of integrable RG models with
different characteristics. The RGK model has a particular value of the
pairing strength $G_{MR}=\frac{1}{L-M+1}$, obtained by Moore and Read
\cite{Moore-Read}, at which the exact ground-state wavefunction is a
pair condensate (PBCS or AGP). From the exact solution, at this point
the $M$ pair energies $E_{\alpha}$ converge to zero transforming the
product of geminals (\ref{PsiRGK}) into the AGP (\ref{PsiMR}). From
the other side, starting with the AGP and making use of the killers we
derived the Moore-Read Hamiltonian (\ref{HMR}) that is contained in
the $\mathcal{G}$ condition, and therefore the v2RDM with the $\PQG$
conditions should provide the exact
solution. Fig. \ref{fig:rgk_energy_diff} gives the numerical proof of
this statement in a highly non-trivial problem. This figure also shows
that the variational method with the $\PQGT$ conditions gives the
exact numerical ground-state energy from weak to strong
pairing. Additional confirmation of the exactness of the $\PQGT$
conditions comes from the canonical gaps in
Fig. \ref{fig:rgk_gap_diff}, which also shows an exact value for $\PQG$
at the Moore-Read point. Similar results for the ground-state energies and
gaps were obtained for the reduced BCS Hamiltonian with equidistant
single-particle levels.  We then tested the robustness of the $\PQGT$
$N$-representability conditions against disorder in the
single-particle levels as in the case of small metallic grains (see
Fig. \ref{fig:random_pairing_energy_diff}). Surprisingly, and even
though the systems are always quantum integrable, the exactness of the
numerical results was lost in the superconducting region
($G>G_c$). This fact might be explained by the complexity of the
ground-state wavefunctions in most of the random instances, as can be
deduced from the distribution of pair energies $E_{\alpha}$ in the
complex plane when the system enters the superconducting
phase. However, relative errors of $10^{-4}$ are still competitive
with DMRG calculations \cite{Sierra1999} for equidistant levels, and with more recent approaches tested in the Richardson model for small size systems \cite{Scuseria, Duguet}.

The exact solvability of these models allowed us to test the v2RDM
method for large systems in order to asses its extensive
properties. Fig. \ref{fig:rgk_energy_various_levels} and
\ref{fig:pairing_energy_various_levels} demonstrate that the high
accuracy of the $\PQGT$ is independent of the system size in the
studied range from $L=10$ to $L=100$.

Before closing, we would like to point out that SU(2) Hamiltonians
encompass the area of quantum magnetism with Heisenberg type
Hamiltonians. The formalism developed in \cite{Alcoba2018} and tested
in this work could be directly applied to the study of spin
systems. Due to the non-perturbative nature of v2RDM, it might be
possible to describe with high accuracy exotic phases and quantum
phase transitions.

\section{Acknowledgement}
A. R. and J. D. acknowledge the financial support of the Spanish
Ministerio de Econom\'ia y Competitividad and the European regional
development fund (FEDER) under Projects
No. FIS2015-63770-P. D. R. A. acknowledges financial support of the
Consejo Nacional de Investigaciones Cient\'ificas y T\'ecnicas under
Grants Nos. PIP 11220130100377CO and 2013-1401PCB, and of the Agencia
Nacional de Promoci\'on Cient\'ifica y Tecnol\'ogica, Argentina under Grant
No. PICT-201-0381. P. C. acknowledges financial support of the Consejo
Nacional de Investigaciones Cient\'ificas y T\'ecnicas under
Grant. No. PIP 11220150100442CO. D. R. A. and P. C. acknowledge
financial support of the Universidad de Buenos Aires under Grant
No. 20020150100157BA.

\bibliographystyle{biochem}
\bibliography{bibi.bib}

\providecommand{\latin}[1]{#1}
\providecommand*\mcitethebibliography{\thebibliography}
\csname @ifundefined\endcsname{endmcitethebibliography}
  {\let\endmcitethebibliography\endthebibliography}{}
\begin{mcitethebibliography}{80}
\providecommand*\natexlab[1]{#1}
\providecommand*\mciteSetBstSublistMode[1]{}
\providecommand*\mciteSetBstMaxWidthForm[2]{}
\providecommand*\mciteBstWouldAddEndPuncttrue
  {\def\EndOfBibitem{\unskip.}}
\providecommand*\mciteBstWouldAddEndPunctfalse
  {\let\EndOfBibitem\relax}
\providecommand*\mciteSetBstMidEndSepPunct[3]{}
\providecommand*\mciteSetBstSublistLabelBeginEnd[3]{}
\providecommand*\EndOfBibitem{}
\mciteSetBstSublistMode{f}
\mciteSetBstMaxWidthForm{subitem}{(\alph{mcitesubitemcount})}
\mciteSetBstSublistLabelBeginEnd
  {\mcitemaxwidthsubitemform\space}
  {\relax}
  {\relax}

\bibitem[Kohn(2003)]{Kohn2000}
Kohn,~W. \emph{Nobel Lectures, Chemistry, 1996-2000}; World Scientific:
  Singapore, 2003; p 213\relax
\mciteBstWouldAddEndPuncttrue
\mciteSetBstMidEndSepPunct{\mcitedefaultmidpunct}
{\mcitedefaultendpunct}{\mcitedefaultseppunct}\relax
\EndOfBibitem
\bibitem[Shavitt and Bartlett(2009)Shavitt, and Bartlett]{Shavitt}
Shavitt,~I.; Bartlett,~R.~J. \emph{Many-Body Methods in Chemistry and Physics:
  MBPT and Coupled-Cluster Theory}; Cambridge Molecular Science; Cambridge
  University Press, 2009\relax
\mciteBstWouldAddEndPuncttrue
\mciteSetBstMidEndSepPunct{\mcitedefaultmidpunct}
{\mcitedefaultendpunct}{\mcitedefaultseppunct}\relax
\EndOfBibitem
\bibitem[Bartlett and Musia\l{}(2007)Bartlett, and Musia\l{}]{Bartlett}
Bartlett,~R.~J.; Musia\l{},~M. Coupled-cluster theory in quantum chemistry.
  \emph{Rev. Mod. Phys.} \textbf{2007}, \emph{79}, 291--352\relax
\mciteBstWouldAddEndPuncttrue
\mciteSetBstMidEndSepPunct{\mcitedefaultmidpunct}
{\mcitedefaultendpunct}{\mcitedefaultseppunct}\relax
\EndOfBibitem
\bibitem[Hagen \latin{et~al.}(2014)Hagen, Papenbrock, Hjorth-Jensen, and
  Dean]{Hagen}
Hagen,~G.; Papenbrock,~T.; Hjorth-Jensen,~M.; Dean,~D.~J. Coupled-cluster
  computations of atomic nuclei. \emph{Reports on Progress in Physics}
  \textbf{2014}, \emph{77}, 096302\relax
\mciteBstWouldAddEndPuncttrue
\mciteSetBstMidEndSepPunct{\mcitedefaultmidpunct}
{\mcitedefaultendpunct}{\mcitedefaultseppunct}\relax
\EndOfBibitem
\bibitem[White(1992)]{White1992}
White,~S.~R. Density matrix formulation for quantum renormalization groups.
  \emph{Phys. Rev. Lett.} \textbf{1992}, \emph{69}, 2863--2866\relax
\mciteBstWouldAddEndPuncttrue
\mciteSetBstMidEndSepPunct{\mcitedefaultmidpunct}
{\mcitedefaultendpunct}{\mcitedefaultseppunct}\relax
\EndOfBibitem
\bibitem[Niggemann \latin{et~al.}(1997)Niggemann, Kl{\"u}mper, and
  Zittartz]{Niggemann1997}
Niggemann,~H.; Kl{\"u}mper,~A.; Zittartz,~J. Quantum phase transition in
  spin-3/2 systems on the hexagonal lattice --- optimum ground state approach.
  \emph{Zeitschrift f{\"u}r Physik B Condensed Matter} \textbf{1997},
  \emph{104}, 103--110\relax
\mciteBstWouldAddEndPuncttrue
\mciteSetBstMidEndSepPunct{\mcitedefaultmidpunct}
{\mcitedefaultendpunct}{\mcitedefaultseppunct}\relax
\EndOfBibitem
\bibitem[Schollw\"ock(2005)]{Schollwock2005}
Schollw\"ock,~U. The density-matrix renormalization group. \emph{Rev. Mod.
  Phys.} \textbf{2005}, \emph{77}, 259--315\relax
\mciteBstWouldAddEndPuncttrue
\mciteSetBstMidEndSepPunct{\mcitedefaultmidpunct}
{\mcitedefaultendpunct}{\mcitedefaultseppunct}\relax
\EndOfBibitem
\bibitem[Vidal(2007)]{Vidal2007}
Vidal,~G. Entanglement Renormalization. \emph{Phys. Rev. Lett.} \textbf{2007},
  \emph{99}, 220405\relax
\mciteBstWouldAddEndPuncttrue
\mciteSetBstMidEndSepPunct{\mcitedefaultmidpunct}
{\mcitedefaultendpunct}{\mcitedefaultseppunct}\relax
\EndOfBibitem
\bibitem[McMillan(1965)]{McMillan1965}
McMillan,~W.~L. Ground State of Liquid ${\mathrm{He}}^{4}$. \emph{Phys. Rev.}
  \textbf{1965}, \emph{138}, A442--A451\relax
\mciteBstWouldAddEndPuncttrue
\mciteSetBstMidEndSepPunct{\mcitedefaultmidpunct}
{\mcitedefaultendpunct}{\mcitedefaultseppunct}\relax
\EndOfBibitem
\bibitem[Ceperley \latin{et~al.}(1977)Ceperley, Chester, and
  Kalos]{Ceperley1977}
Ceperley,~D.; Chester,~G.~V.; Kalos,~M.~H. Monte Carlo simulation of a
  many-fermion study. \emph{Phys. Rev. B} \textbf{1977}, \emph{16},
  3081--3099\relax
\mciteBstWouldAddEndPuncttrue
\mciteSetBstMidEndSepPunct{\mcitedefaultmidpunct}
{\mcitedefaultendpunct}{\mcitedefaultseppunct}\relax
\EndOfBibitem
\bibitem[Mezzacapo \latin{et~al.}(2009)Mezzacapo, Schuch, Boninsegni, and
  Cirac]{Mezzacapo2009}
Mezzacapo,~F.; Schuch,~N.; Boninsegni,~M.; Cirac,~J.~I. Ground-state properties
  of quantum many-body systems: entangled-plaquette states and variational
  Monte Carlo. \emph{New Journal of Physics} \textbf{2009}, \emph{11},
  083026\relax
\mciteBstWouldAddEndPuncttrue
\mciteSetBstMidEndSepPunct{\mcitedefaultmidpunct}
{\mcitedefaultendpunct}{\mcitedefaultseppunct}\relax
\EndOfBibitem
\bibitem[Changlani \latin{et~al.}(2009)Changlani, Kinder, Umrigar, and
  Chan]{Changlani2009}
Changlani,~H.~J.; Kinder,~J.~M.; Umrigar,~C.~J.; Chan,~G. K.-L. Approximating
  strongly correlated wave functions with correlator product states.
  \emph{Phys. Rev. B} \textbf{2009}, \emph{80}, 245116\relax
\mciteBstWouldAddEndPuncttrue
\mciteSetBstMidEndSepPunct{\mcitedefaultmidpunct}
{\mcitedefaultendpunct}{\mcitedefaultseppunct}\relax
\EndOfBibitem
\bibitem[Suzuki \latin{et~al.}(1977)Suzuki, Miyashita, and Kuroda]{Suzuki1977}
Suzuki,~M.; Miyashita,~S.; Kuroda,~A. Monte Carlo Simulation of Quantum Spin
  Systems. I. \emph{Progress of Theoretical Physics} \textbf{1977}, \emph{58},
  1377--1387\relax
\mciteBstWouldAddEndPuncttrue
\mciteSetBstMidEndSepPunct{\mcitedefaultmidpunct}
{\mcitedefaultendpunct}{\mcitedefaultseppunct}\relax
\EndOfBibitem
\bibitem[Prokof'ev \latin{et~al.}(1996)Prokof'ev, Svistunov, and
  Tupitsyn]{Prokofev1996}
Prokof'ev,~N.~V.; Svistunov,~B.~V.; Tupitsyn,~I.~S. Exact quantum Monte Carlo
  process for the statistics of discrete systems. \emph{Journal of Experimental
  and Theoretical Physics Letters} \textbf{1996}, \emph{64}, 911--916\relax
\mciteBstWouldAddEndPuncttrue
\mciteSetBstMidEndSepPunct{\mcitedefaultmidpunct}
{\mcitedefaultendpunct}{\mcitedefaultseppunct}\relax
\EndOfBibitem
\bibitem[Sylju\aa{}sen and Sandvik(2002)Sylju\aa{}sen, and
  Sandvik]{Syljuasen2002}
Sylju\aa{}sen,~O.~F.; Sandvik,~A.~W. Quantum Monte Carlo with directed loops.
  \emph{Phys. Rev. E} \textbf{2002}, \emph{66}, 046701\relax
\mciteBstWouldAddEndPuncttrue
\mciteSetBstMidEndSepPunct{\mcitedefaultmidpunct}
{\mcitedefaultendpunct}{\mcitedefaultseppunct}\relax
\EndOfBibitem
\bibitem[Alet \latin{et~al.}(2005)Alet, Wessel, and Troyer]{Alet2005}
Alet,~F.; Wessel,~S.; Troyer,~M. Generalized directed loop method for quantum
  Monte Carlo simulations. \emph{Phys. Rev. E} \textbf{2005}, \emph{71},
  036706\relax
\mciteBstWouldAddEndPuncttrue
\mciteSetBstMidEndSepPunct{\mcitedefaultmidpunct}
{\mcitedefaultendpunct}{\mcitedefaultseppunct}\relax
\EndOfBibitem
\bibitem[Motta \latin{et~al.}(2017)Motta, Ceperley, Chan, Gomez, Gull, Guo,
  Jim\'enez-Hoyos, Lan, Li, Ma, Millis, Prokof'ev, Ray, Scuseria, Sorella,
  Stoudenmire, Sun, Tupitsyn, White, Zgid, and Zhang]{Motta}
Motta,~M.; Ceperley,~D.~M.; Chan,~G. K.-L.; Gomez,~J.~A.; Gull,~E.; Guo,~S.;
  Jim\'enez-Hoyos,~C.~A.; Lan,~T.~N.; Li,~J.; Ma,~F.; Millis,~A.~J.;
  Prokof'ev,~N.~V.; Ray,~U.; Scuseria,~G.~E.; Sorella,~S.; Stoudenmire,~E.~M.;
  Sun,~Q.; Tupitsyn,~I.~S.; White,~S.~R.; Zgid,~D.; Zhang,~S. Towards the
  Solution of the Many-Electron Problem in Real Materials: Equation of State of
  the Hydrogen Chain with State-of-the-Art Many-Body Methods. \emph{Phys. Rev.
  X} \textbf{2017}, \emph{7}, 031059\relax
\mciteBstWouldAddEndPuncttrue
\mciteSetBstMidEndSepPunct{\mcitedefaultmidpunct}
{\mcitedefaultendpunct}{\mcitedefaultseppunct}\relax
\EndOfBibitem
\bibitem[Husimi(1940)]{Husimi1940}
Husimi,~K. Some Formal Properties of the Density Matrix. \emph{Proceedings of
  the Physico-Mathematical Society of Japan. 3rd Series} \textbf{1940},
  \emph{22}, 264--314\relax
\mciteBstWouldAddEndPuncttrue
\mciteSetBstMidEndSepPunct{\mcitedefaultmidpunct}
{\mcitedefaultendpunct}{\mcitedefaultseppunct}\relax
\EndOfBibitem
\bibitem[L\"owdin(1955)]{Lowdin1955}
L\"owdin,~P.-O. Quantum Theory of Many-Particle Systems. I. Physical
  Interpretations by Means of Density Matrices, Natural Spin-Orbitals, and
  Convergence Problems in the Method of Configurational Interaction.
  \emph{Phys. Rev.} \textbf{1955}, \emph{97}, 1474--1489\relax
\mciteBstWouldAddEndPuncttrue
\mciteSetBstMidEndSepPunct{\mcitedefaultmidpunct}
{\mcitedefaultendpunct}{\mcitedefaultseppunct}\relax
\EndOfBibitem
\bibitem[Mayer(1955)]{Mayer1955}
Mayer,~J.~E. Electron Correlation. \emph{Phys. Rev.} \textbf{1955}, \emph{100},
  1579--1586\relax
\mciteBstWouldAddEndPuncttrue
\mciteSetBstMidEndSepPunct{\mcitedefaultmidpunct}
{\mcitedefaultendpunct}{\mcitedefaultseppunct}\relax
\EndOfBibitem
\bibitem[Tredgold(1957)]{Tredgold1957}
Tredgold,~R.~H. Density Matrix and the Many-Body Problem. \emph{Phys. Rev.}
  \textbf{1957}, \emph{105}, 1421--1423\relax
\mciteBstWouldAddEndPuncttrue
\mciteSetBstMidEndSepPunct{\mcitedefaultmidpunct}
{\mcitedefaultendpunct}{\mcitedefaultseppunct}\relax
\EndOfBibitem
\bibitem[Coleman(1963)]{Coleman1963}
Coleman,~A.~J. Structure of Fermion Density Matrices. \emph{Rev. Mod. Phys.}
  \textbf{1963}, \emph{35}, 668--686\relax
\mciteBstWouldAddEndPuncttrue
\mciteSetBstMidEndSepPunct{\mcitedefaultmidpunct}
{\mcitedefaultendpunct}{\mcitedefaultseppunct}\relax
\EndOfBibitem
\bibitem[Liu \latin{et~al.}(2007)Liu, Christandl, and Verstraete]{Liu2007}
Liu,~Y.-K.; Christandl,~M.; Verstraete,~F. Quantum Computational Complexity of
  the $N$-Representability Problem: QMA Complete. \emph{Phys. Rev. Lett.}
  \textbf{2007}, \emph{98}, 110503\relax
\mciteBstWouldAddEndPuncttrue
\mciteSetBstMidEndSepPunct{\mcitedefaultmidpunct}
{\mcitedefaultendpunct}{\mcitedefaultseppunct}\relax
\EndOfBibitem
\bibitem[Garrod \latin{et~al.}(1975)Garrod, Mihailovi\'c, and
  Rosina]{Garrod1975}
Garrod,~C.; Mihailovi\'c,~M.~V.; Rosina,~M. The variational approach to the
  two-body density matrix. \emph{Journal of Mathematical Physics}
  \textbf{1975}, \emph{16}, 868--874\relax
\mciteBstWouldAddEndPuncttrue
\mciteSetBstMidEndSepPunct{\mcitedefaultmidpunct}
{\mcitedefaultendpunct}{\mcitedefaultseppunct}\relax
\EndOfBibitem
\bibitem[Nakata \latin{et~al.}(2001)Nakata, Nakatsuji, Ehara, Fukuda, Nakata,
  and Fujisawa]{Nakata2001}
Nakata,~M.; Nakatsuji,~H.; Ehara,~M.; Fukuda,~M.; Nakata,~K.; Fujisawa,~K.
  Variational calculations of fermion second-order reduced density matrices by
  semidefinite programming algorithm. \emph{The Journal of Chemical Physics}
  \textbf{2001}, \emph{114}, 8282--8292\relax
\mciteBstWouldAddEndPuncttrue
\mciteSetBstMidEndSepPunct{\mcitedefaultmidpunct}
{\mcitedefaultendpunct}{\mcitedefaultseppunct}\relax
\EndOfBibitem
\bibitem[Mazziotti(2002)]{Mazziotti2002}
Mazziotti,~D.~A. Variational minimization of atomic and molecular ground-state
  energies via the two-particle reduced density matrix. \emph{Phys. Rev. A}
  \textbf{2002}, \emph{65}, 062511\relax
\mciteBstWouldAddEndPuncttrue
\mciteSetBstMidEndSepPunct{\mcitedefaultmidpunct}
{\mcitedefaultendpunct}{\mcitedefaultseppunct}\relax
\EndOfBibitem
\bibitem[Zhao \latin{et~al.}(2004)Zhao, Braams, Fukuda, Overton, and
  Percus]{Zhao2004}
Zhao,~Z.; Braams,~B.~J.; Fukuda,~M.; Overton,~M.~L.; Percus,~J.~K. The reduced
  density matrix method for electronic structure calculations and the role of
  three-index representability conditions. \emph{The Journal of Chemical
  Physics} \textbf{2004}, \emph{120}, 2095--2104\relax
\mciteBstWouldAddEndPuncttrue
\mciteSetBstMidEndSepPunct{\mcitedefaultmidpunct}
{\mcitedefaultendpunct}{\mcitedefaultseppunct}\relax
\EndOfBibitem
\bibitem[Mihailovi\'c and Rosina(1975)Mihailovi\'c, and Rosina]{Rosina}
Mihailovi\'c,~M.; Rosina,~M. The variational approach to the density matrix for
  light nuclei. \emph{Nuclear Physics A} \textbf{1975}, \emph{237}, 221 --
  228\relax
\mciteBstWouldAddEndPuncttrue
\mciteSetBstMidEndSepPunct{\mcitedefaultmidpunct}
{\mcitedefaultendpunct}{\mcitedefaultseppunct}\relax
\EndOfBibitem
\bibitem[Verstichel \latin{et~al.}(2011)Verstichel, van Aggelen, Neck,
  Bultinck, and Baerdemacker]{Exact}
Verstichel,~B.; van Aggelen,~H.; Neck,~D.~V.; Bultinck,~P.; Baerdemacker,~S.~D.
  A primal–dual semidefinite programming algorithm tailored to the
  variational determination of the two-body density matrix. \emph{Computer
  Physics Communications} \textbf{2011}, \emph{182}, 1235 -- 1244\relax
\mciteBstWouldAddEndPuncttrue
\mciteSetBstMidEndSepPunct{\mcitedefaultmidpunct}
{\mcitedefaultendpunct}{\mcitedefaultseppunct}\relax
\EndOfBibitem
\bibitem[Hammond and Mazziotti(2006)Hammond, and Mazziotti]{Hubbard1}
Hammond,~J.~R.; Mazziotti,~D.~A. Variational reduced-density-matrix calculation
  of the one-dimensional Hubbard model. \emph{Phys. Rev. A} \textbf{2006},
  \emph{73}, 062505\relax
\mciteBstWouldAddEndPuncttrue
\mciteSetBstMidEndSepPunct{\mcitedefaultmidpunct}
{\mcitedefaultendpunct}{\mcitedefaultseppunct}\relax
\EndOfBibitem
\bibitem[Verstichel \latin{et~al.}(2013)Verstichel, van Aggelen, Poelmans,
  Wouters, and Neck]{Hubbard2}
Verstichel,~B.; van Aggelen,~H.; Poelmans,~W.; Wouters,~S.; Neck,~D.~V.
  Extensive v2DM study of the one-dimensional Hubbard model for large lattice
  sizes: Exploiting translational invariance and parity. \emph{Computational
  and Theoretical Chemistry} \textbf{2013}, \emph{1003}, 12 -- 21, Reduced
  Density Matrices: A Simpler Approach to Many-Electron Problems?\relax
\mciteBstWouldAddEndPunctfalse
\mciteSetBstMidEndSepPunct{\mcitedefaultmidpunct}
{}{\mcitedefaultseppunct}\relax
\EndOfBibitem
\bibitem[Anderson \latin{et~al.}(2013)Anderson, Nakata, Igarashi, Fujisawa, and
  Yamashita]{Hubbard3}
Anderson,~J.~S.; Nakata,~M.; Igarashi,~R.; Fujisawa,~K.; Yamashita,~M. The
  second-order reduced density matrix method and the two-dimensional Hubbard
  model. \emph{Computational and Theoretical Chemistry} \textbf{2013},
  \emph{1003}, 22 -- 27, Reduced Density Matrices: A Simpler Approach to
  Many-Electron Problems?\relax
\mciteBstWouldAddEndPunctfalse
\mciteSetBstMidEndSepPunct{\mcitedefaultmidpunct}
{}{\mcitedefaultseppunct}\relax
\EndOfBibitem
\bibitem[Talmi(1993)]{Talmi}
Talmi,~I. \emph{Simple models of complex nuclei}; Chur, Switzerland ;
  Langhorne, Pa., U.S.A.: Harwood Academic Publishers, 1993\relax
\mciteBstWouldAddEndPuncttrue
\mciteSetBstMidEndSepPunct{\mcitedefaultmidpunct}
{\mcitedefaultendpunct}{\mcitedefaultseppunct}\relax
\EndOfBibitem
\bibitem[Bytautas \latin{et~al.}(2011)Bytautas, Henderson, Jiménez-Hoyos,
  Ellis, and Scuseria]{Bitautas}
Bytautas,~L.; Henderson,~T.~M.; Jiménez-Hoyos,~C.~A.; Ellis,~J.~K.;
  Scuseria,~G.~E. Seniority and orbital symmetry as tools for establishing a
  full configuration interaction hierarchy. \emph{The Journal of Chemical
  Physics} \textbf{2011}, \emph{135}, 044119\relax
\mciteBstWouldAddEndPuncttrue
\mciteSetBstMidEndSepPunct{\mcitedefaultmidpunct}
{\mcitedefaultendpunct}{\mcitedefaultseppunct}\relax
\EndOfBibitem
\bibitem[Alcoba \latin{et~al.}(2013)Alcoba, Torre, Lain, Massaccesi, and
  Oña]{Alcoba2013}
Alcoba,~D.~R.; Torre,~A.; Lain,~L.; Massaccesi,~G.~E.; Oña,~O.~B. Seniority
  number in spin-adapted spaces and compactness of configuration interaction
  wave functions. \emph{The Journal of Chemical Physics} \textbf{2013},
  \emph{139}, 084103\relax
\mciteBstWouldAddEndPuncttrue
\mciteSetBstMidEndSepPunct{\mcitedefaultmidpunct}
{\mcitedefaultendpunct}{\mcitedefaultseppunct}\relax
\EndOfBibitem
\bibitem[Limacher \latin{et~al.}(2013)Limacher, Ayers, Johnson,
  De~Baerdemacker, Van~Neck, and Bultinck]{Limacher2013}
Limacher,~P.~A.; Ayers,~P.~W.; Johnson,~P.~A.; De~Baerdemacker,~S.;
  Van~Neck,~D.; Bultinck,~P. A New Mean-Field Method Suitable for Strongly
  Correlated Electrons: Computationally Facile Antisymmetric Products of
  Nonorthogonal Geminals. \emph{Journal of Chemical Theory and Computation}
  \textbf{2013}, \emph{9}, 1394--1401\relax
\mciteBstWouldAddEndPuncttrue
\mciteSetBstMidEndSepPunct{\mcitedefaultmidpunct}
{\mcitedefaultendpunct}{\mcitedefaultseppunct}\relax
\EndOfBibitem
\bibitem[Alcoba \latin{et~al.}(2014)Alcoba, Torre, Lain, Massaccesi, and
  Oña]{Alcoba2014}
Alcoba,~D.~R.; Torre,~A.; Lain,~L.; Massaccesi,~G.~E.; Oña,~O.~B.
  Configuration interaction wave functions: A seniority number approach.
  \emph{The Journal of Chemical Physics} \textbf{2014}, \emph{140},
  234103\relax
\mciteBstWouldAddEndPuncttrue
\mciteSetBstMidEndSepPunct{\mcitedefaultmidpunct}
{\mcitedefaultendpunct}{\mcitedefaultseppunct}\relax
\EndOfBibitem
\bibitem[Weinhold and Wilson(1967)Weinhold, and Wilson]{Weinhold1967}
Weinhold,~F.; Wilson,~E.~B. Reduced Density Matrices of Atoms and Molecules. I.
  The 2 Matrix of Double‐Occupancy, Configuration‐Interaction Wavefunctions
  for Singlet States. \emph{The Journal of Chemical Physics} \textbf{1967},
  \emph{46}, 2752--2758\relax
\mciteBstWouldAddEndPuncttrue
\mciteSetBstMidEndSepPunct{\mcitedefaultmidpunct}
{\mcitedefaultendpunct}{\mcitedefaultseppunct}\relax
\EndOfBibitem
\bibitem[Weinhold and Wilson(1967)Weinhold, and Wilson]{Weinhold1967b}
Weinhold,~F.; Wilson,~E.~B. Reduced Density Matrices of Atoms and Molecules.
  II. On the N‐Representability Problem. \emph{The Journal of Chemical
  Physics} \textbf{1967}, \emph{47}, 2298--2311\relax
\mciteBstWouldAddEndPuncttrue
\mciteSetBstMidEndSepPunct{\mcitedefaultmidpunct}
{\mcitedefaultendpunct}{\mcitedefaultseppunct}\relax
\EndOfBibitem
\bibitem[Poelmans \latin{et~al.}(2015)Poelmans, Van~Raemdonck, Verstichel,
  De~Baerdemacker, Torre, Lain, Massaccesi, Alcoba, Bultinck, and
  Van~Neck]{Poelmans2015}
Poelmans,~W.; Van~Raemdonck,~M.; Verstichel,~B.; De~Baerdemacker,~S.;
  Torre,~A.; Lain,~L.; Massaccesi,~G.~E.; Alcoba,~D.~R.; Bultinck,~P.;
  Van~Neck,~D. Variational Optimization of the Second-Order Density Matrix
  Corresponding to a Seniority-Zero Configuration Interaction Wave Function.
  \emph{Journal of Chemical Theory and Computation} \textbf{2015}, \emph{11},
  4064--4076\relax
\mciteBstWouldAddEndPuncttrue
\mciteSetBstMidEndSepPunct{\mcitedefaultmidpunct}
{\mcitedefaultendpunct}{\mcitedefaultseppunct}\relax
\EndOfBibitem
\bibitem[Head-Marsden and Mazziotti(2017)Head-Marsden, and
  Mazziotti]{Head-Marsden2017}
Head-Marsden,~K.; Mazziotti,~D.~A. Pair 2-electron reduced density matrix
  theory using localized orbitals. \emph{The Journal of Chemical Physics}
  \textbf{2017}, \emph{147}, 084101\relax
\mciteBstWouldAddEndPuncttrue
\mciteSetBstMidEndSepPunct{\mcitedefaultmidpunct}
{\mcitedefaultendpunct}{\mcitedefaultseppunct}\relax
\EndOfBibitem
\bibitem[Alcoba \latin{et~al.}(2018)Alcoba, Torre, Lain, Massaccesi, Oña,
  Honoré, Poelmans, Neck, Bultinck, and Baerdemacker]{Alcoba2018}
Alcoba,~D.~R.; Torre,~A.; Lain,~L.; Massaccesi,~G.~E.; Oña,~O.~B.;
  Honoré,~E.~M.; Poelmans,~W.; Neck,~D.~V.; Bultinck,~P.; Baerdemacker,~S.~D.
  Direct variational determination of the two-electron reduced density matrix
  for doubly occupied-configuration-interaction wave functions: The influence
  of three-index N-representability conditions. \emph{The Journal of Chemical
  Physics} \textbf{2018}, \emph{148}, 024105\relax
\mciteBstWouldAddEndPuncttrue
\mciteSetBstMidEndSepPunct{\mcitedefaultmidpunct}
{\mcitedefaultendpunct}{\mcitedefaultseppunct}\relax
\EndOfBibitem
\bibitem[Poelmans(2015)]{Poelmans2015Thesis}
Poelmans,~W. Variational determination of the two-particle density matrix: The
  case of doubly-occupied space. Ph.D.\ thesis, Ghent University, 2015\relax
\mciteBstWouldAddEndPuncttrue
\mciteSetBstMidEndSepPunct{\mcitedefaultmidpunct}
{\mcitedefaultendpunct}{\mcitedefaultseppunct}\relax
\EndOfBibitem
\bibitem[Dukelsky \latin{et~al.}(2001)Dukelsky, Esebbag, and
  Schuck]{Dukelsky2001}
Dukelsky,~J.; Esebbag,~C.; Schuck,~P. Class of Exactly Solvable Pairing Models.
  \emph{Phys. Rev. Lett.} \textbf{2001}, \emph{87}, 066403\relax
\mciteBstWouldAddEndPuncttrue
\mciteSetBstMidEndSepPunct{\mcitedefaultmidpunct}
{\mcitedefaultendpunct}{\mcitedefaultseppunct}\relax
\EndOfBibitem
\bibitem[Dukelsky \latin{et~al.}(2004)Dukelsky, Pittel, and
  Sierra]{Dukelsky2004}
Dukelsky,~J.; Pittel,~S.; Sierra,~G. Colloquium: Exactly solvable
  Richardson-Gaudin models for many-body quantum systems. \emph{Rev. Mod.
  Phys.} \textbf{2004}, \emph{76}, 643--662\relax
\mciteBstWouldAddEndPuncttrue
\mciteSetBstMidEndSepPunct{\mcitedefaultmidpunct}
{\mcitedefaultendpunct}{\mcitedefaultseppunct}\relax
\EndOfBibitem
\bibitem[Ortiz \latin{et~al.}(2005)Ortiz, Somma, Dukelsky, and Rombouts]{CCT}
Ortiz,~G.; Somma,~R.; Dukelsky,~J.; Rombouts,~S. Exactly-solvable models
  derived from a generalized Gaudin algebra. \emph{Nuclear Physics B}
  \textbf{2005}, \emph{707}, 421 -- 457\relax
\mciteBstWouldAddEndPuncttrue
\mciteSetBstMidEndSepPunct{\mcitedefaultmidpunct}
{\mcitedefaultendpunct}{\mcitedefaultseppunct}\relax
\EndOfBibitem
\bibitem[Ortiz \latin{et~al.}(2014)Ortiz, Dukelsky, Cobanera, Esebbag, and
  Beenakker]{Ortiz}
Ortiz,~G.; Dukelsky,~J.; Cobanera,~E.; Esebbag,~C.; Beenakker,~C. Many-Body
  Characterization of Particle-Conserving Topological Superfluids. \emph{Phys.
  Rev. Lett.} \textbf{2014}, \emph{113}, 267002\relax
\mciteBstWouldAddEndPuncttrue
\mciteSetBstMidEndSepPunct{\mcitedefaultmidpunct}
{\mcitedefaultendpunct}{\mcitedefaultseppunct}\relax
\EndOfBibitem
\bibitem[Richardson(1966)]{Richardson}
Richardson,~R.~W. Numerical Study of the 8-32-Particle Eigenstates of the
  Pairing Hamiltonian. \emph{Phys. Rev.} \textbf{1966}, \emph{141},
  949--956\relax
\mciteBstWouldAddEndPuncttrue
\mciteSetBstMidEndSepPunct{\mcitedefaultmidpunct}
{\mcitedefaultendpunct}{\mcitedefaultseppunct}\relax
\EndOfBibitem
\bibitem[Sierra \latin{et~al.}(2000)Sierra, Dukelsky, Dussel, von Delft, and
  Braun]{Delft}
Sierra,~G.; Dukelsky,~J.; Dussel,~G.~G.; von Delft,~J.; Braun,~F. Exact study
  of the effect of level statistics in ultrasmall superconducting grains.
  \emph{Phys. Rev. B} \textbf{2000}, \emph{61}, R11890--R11893\relax
\mciteBstWouldAddEndPuncttrue
\mciteSetBstMidEndSepPunct{\mcitedefaultmidpunct}
{\mcitedefaultendpunct}{\mcitedefaultseppunct}\relax
\EndOfBibitem
\bibitem[Jorgensen(1981)]{Jorgensen1981}
Jorgensen,~P. \emph{Second Quantization-Based Methods in Quantum Chemistry};
  Academic Press: New York, 1981\relax
\mciteBstWouldAddEndPuncttrue
\mciteSetBstMidEndSepPunct{\mcitedefaultmidpunct}
{\mcitedefaultendpunct}{\mcitedefaultseppunct}\relax
\EndOfBibitem
\bibitem[Coleman(2000)]{Coleman2000}
Coleman,~A.~J. In \emph{Many-electron densities and reduced density matrices},
  1st ed.; Cioslowski,~J., Ed.; Springer Science+Business Media: New York,
  2000; p~1\relax
\mciteBstWouldAddEndPuncttrue
\mciteSetBstMidEndSepPunct{\mcitedefaultmidpunct}
{\mcitedefaultendpunct}{\mcitedefaultseppunct}\relax
\EndOfBibitem
\bibitem[Garrod and Percus(1964)Garrod, and Percus]{Garrod1964}
Garrod,~C.; Percus,~J.~K. Reduction of the N‐Particle Variational Problem.
  \emph{Journal of Mathematical Physics} \textbf{1964}, \emph{5},
  1756--1776\relax
\mciteBstWouldAddEndPuncttrue
\mciteSetBstMidEndSepPunct{\mcitedefaultmidpunct}
{\mcitedefaultendpunct}{\mcitedefaultseppunct}\relax
\EndOfBibitem
\bibitem[Kummer(1967)]{Kummer1967}
Kummer,~H. n‐Representability Problem for Reduced Density Matrices.
  \emph{Journal of Mathematical Physics} \textbf{1967}, \emph{8},
  2063--2081\relax
\mciteBstWouldAddEndPuncttrue
\mciteSetBstMidEndSepPunct{\mcitedefaultmidpunct}
{\mcitedefaultendpunct}{\mcitedefaultseppunct}\relax
\EndOfBibitem
\bibitem[Coleman(1974)]{Coleman1974}
Coleman,~A.~J. In \emph{Reduced Density Operators with Applications to Physical
  and Chemical Systems - II, Queen’s Papers on Pure and Applied Mathematics};
  Erdahl,~R.~M., Ed.; Queens University: Kingston, Ontario, 1974; p~2\relax
\mciteBstWouldAddEndPuncttrue
\mciteSetBstMidEndSepPunct{\mcitedefaultmidpunct}
{\mcitedefaultendpunct}{\mcitedefaultseppunct}\relax
\EndOfBibitem
\bibitem[Coleman and Yukalov(2000)Coleman, and Yukalov]{Coleman2000b}
Coleman,~A.~J.; Yukalov,~V.~I. \emph{Reduced Density Matrices: Coulson’s
  Challange}; Springer-Verlag: New York, 2000\relax
\mciteBstWouldAddEndPuncttrue
\mciteSetBstMidEndSepPunct{\mcitedefaultmidpunct}
{\mcitedefaultendpunct}{\mcitedefaultseppunct}\relax
\EndOfBibitem
\bibitem[Erdahl and Rosina(1974)Erdahl, and Rosina]{Erdhal1974}
Erdahl,~R.~M.; Rosina,~M. In \emph{Reduced Density Operators with Applications
  to Physical and Chemical Systems - II, Queen’s Papers in Pure and Applied
  Mathematics}; Erdahl,~R.~M., Ed.; Queens University: Kingston, Ontario, 1974;
  p~36\relax
\mciteBstWouldAddEndPuncttrue
\mciteSetBstMidEndSepPunct{\mcitedefaultmidpunct}
{\mcitedefaultendpunct}{\mcitedefaultseppunct}\relax
\EndOfBibitem
\bibitem[Erdahl()]{Erdhal1978}
Erdahl,~R.~M. Representability. \emph{International Journal of Quantum
  Chemistry} \emph{13}, 697--718\relax
\mciteBstWouldAddEndPuncttrue
\mciteSetBstMidEndSepPunct{\mcitedefaultmidpunct}
{\mcitedefaultendpunct}{\mcitedefaultseppunct}\relax
\EndOfBibitem
\bibitem[Mazziotti(2005)]{Mazziotti2005}
Mazziotti,~D.~A. Variational two-electron reduced density matrix theory for
  many-electron atoms and molecules: Implementation of the spin- and
  symmetry-adapted ${T}_{2}$ condition through first-order semidefinite
  programming. \emph{Phys. Rev. A} \textbf{2005}, \emph{72}, 032510\relax
\mciteBstWouldAddEndPuncttrue
\mciteSetBstMidEndSepPunct{\mcitedefaultmidpunct}
{\mcitedefaultendpunct}{\mcitedefaultseppunct}\relax
\EndOfBibitem
\bibitem[Nesterov and Nemirovskii(1994)Nesterov, and Nemirovskii]{Nesterov1993}
Nesterov,~Y.; Nemirovskii,~A. \emph{Interior-Point Polynomial Algorithms in
  Convex Programming}; Society for Industrial and Applied Mathematics,
  1994\relax
\mciteBstWouldAddEndPuncttrue
\mciteSetBstMidEndSepPunct{\mcitedefaultmidpunct}
{\mcitedefaultendpunct}{\mcitedefaultseppunct}\relax
\EndOfBibitem
\bibitem[Vandenberghe and Boyd(1996)Vandenberghe, and Boyd]{Vandenberghe1996}
Vandenberghe,~L.; Boyd,~S. Semidefinite Programming. \emph{SIAM Review}
  \textbf{1996}, \emph{38}, 49--95\relax
\mciteBstWouldAddEndPuncttrue
\mciteSetBstMidEndSepPunct{\mcitedefaultmidpunct}
{\mcitedefaultendpunct}{\mcitedefaultseppunct}\relax
\EndOfBibitem
\bibitem[Wright(1997)]{Wright1997}
Wright,~S. \emph{Primal-Dual Interior-Point Methods}; Society for Industrial
  and Applied Mathematics, 1997\relax
\mciteBstWouldAddEndPuncttrue
\mciteSetBstMidEndSepPunct{\mcitedefaultmidpunct}
{\mcitedefaultendpunct}{\mcitedefaultseppunct}\relax
\EndOfBibitem
\bibitem[Wright(200)]{Wolkowicz2000}
Wright,~S. In \emph{Handbook of Semidefinite Programming}, 1st ed.;
  Wolkowicz,~H., Saigal,~R., Vandenberghe,~L., Eds.; International Series in
  Operations Research \& Management Science; Springer US, 200; Vol.~27\relax
\mciteBstWouldAddEndPuncttrue
\mciteSetBstMidEndSepPunct{\mcitedefaultmidpunct}
{\mcitedefaultendpunct}{\mcitedefaultseppunct}\relax
\EndOfBibitem
\bibitem[Yamashita \latin{et~al.}(2011)Yamashita, Fujisawa, Fukuda, Kobayashi,
  and Maho~Nakata]{sdpa-family}
Yamashita,~M.; Fujisawa,~K.; Fukuda,~M.; Kobayashi,~K.,~K.~Nakata;
  Maho~Nakata,~M. In \emph{Semidefinite, Cone and Polynomial Optimization};
  Anjos,~M.~F., Lasserre,~J.~B., Eds.; Springer: New York, 2011; p 687\relax
\mciteBstWouldAddEndPuncttrue
\mciteSetBstMidEndSepPunct{\mcitedefaultmidpunct}
{\mcitedefaultendpunct}{\mcitedefaultseppunct}\relax
\EndOfBibitem
\bibitem[Yamashita \latin{et~al.}(2010)Yamashita, Fujisawa, Nakata, Nakata,
  Fukuda, Kobayashi, and Goto]{sdpa-v7}
Yamashita,~M.; Fujisawa,~K.; Nakata,~K.; Nakata,~M.; Fukuda,~M.; Kobayashi,~K.;
  Goto,~K. A high-performance software package for semidefinite programs: SDPA
  7. \textbf{2010}, \relax
\mciteBstWouldAddEndPunctfalse
\mciteSetBstMidEndSepPunct{\mcitedefaultmidpunct}
{}{\mcitedefaultseppunct}\relax
\EndOfBibitem
\bibitem[Nakata \latin{et~al.}(2008)Nakata, Braams, Fujisawa, Fukuda, Percus,
  Yamashita, and Zhao]{Nakata2008}
Nakata,~M.; Braams,~B.~J.; Fujisawa,~K.; Fukuda,~M.; Percus,~J.~K.;
  Yamashita,~M.; Zhao,~Z. Variational calculation of second-order reduced
  density matrices by strong N-representability conditions and an accurate
  semidefinite programming solver. \emph{The Journal of Chemical Physics}
  \textbf{2008}, \emph{128}, 164113\relax
\mciteBstWouldAddEndPuncttrue
\mciteSetBstMidEndSepPunct{\mcitedefaultmidpunct}
{\mcitedefaultendpunct}{\mcitedefaultseppunct}\relax
\EndOfBibitem
\bibitem[Claeys \latin{et~al.}(2015)Claeys, De~Baerdemacker, Van~Raemdonck, and
  Van~Neck]{Pieter2015}
Claeys,~P.~W.; De~Baerdemacker,~S.; Van~Raemdonck,~M.; Van~Neck,~D.
  Eigenvalue-based method and form-factor determinant representations for
  integrable XXZ Richardson-Gaudin models. \emph{Phys. Rev. B} \textbf{2015},
  \emph{91}, 155102\relax
\mciteBstWouldAddEndPuncttrue
\mciteSetBstMidEndSepPunct{\mcitedefaultmidpunct}
{\mcitedefaultendpunct}{\mcitedefaultseppunct}\relax
\EndOfBibitem
\bibitem[Kitaev(2001)]{Kitaev}
Kitaev,~A.~Y. Unpaired Majorana fermions in quantum wires.
  \emph{Physics-Uspekhi} \textbf{2001}, \emph{44}, 131\relax
\mciteBstWouldAddEndPuncttrue
\mciteSetBstMidEndSepPunct{\mcitedefaultmidpunct}
{\mcitedefaultendpunct}{\mcitedefaultseppunct}\relax
\EndOfBibitem
\bibitem[Iba\~nez \latin{et~al.}(2009)Iba\~nez, Links, Sierra, and
  Zhao]{Ibanes}
Iba\~nez,~M.; Links,~J.; Sierra,~G.; Zhao,~S.-Y. Exactly solvable pairing model
  for superconductors with ${p}_{x}+i{p}_{y}$-wave symmetry. \emph{Phys. Rev.
  B} \textbf{2009}, \emph{79}, 180501\relax
\mciteBstWouldAddEndPuncttrue
\mciteSetBstMidEndSepPunct{\mcitedefaultmidpunct}
{\mcitedefaultendpunct}{\mcitedefaultseppunct}\relax
\EndOfBibitem
\bibitem[Rombouts \latin{et~al.}(2010)Rombouts, Dukelsky, and Ortiz]{Rombouts}
Rombouts,~S. M.~A.; Dukelsky,~J.; Ortiz,~G. Quantum phase diagram of the
  integrable ${p}_{x}+i{p}_{y}$ fermionic superfluid. \emph{Phys. Rev. B}
  \textbf{2010}, \emph{82}, 224510\relax
\mciteBstWouldAddEndPuncttrue
\mciteSetBstMidEndSepPunct{\mcitedefaultmidpunct}
{\mcitedefaultendpunct}{\mcitedefaultseppunct}\relax
\EndOfBibitem
\bibitem[Van~Raemdonck \latin{et~al.}(2014)Van~Raemdonck, De~Baerdemacker, and
  Van~Neck]{Neck}
Van~Raemdonck,~M.; De~Baerdemacker,~S.; Van~Neck,~D. Exact solution of the
  ${p}_{x}+i{p}_{y}$ pairing Hamiltonian by deforming the pairing algebra.
  \emph{Phys. Rev. B} \textbf{2014}, \emph{89}, 155136\relax
\mciteBstWouldAddEndPuncttrue
\mciteSetBstMidEndSepPunct{\mcitedefaultmidpunct}
{\mcitedefaultendpunct}{\mcitedefaultseppunct}\relax
\EndOfBibitem
\bibitem[Moore and Read(1991)Moore, and Read]{Moore-Read}
Moore,~G.; Read,~N. Nonabelions in the fractional quantum hall effect.
  \emph{Nuclear Physics B} \textbf{1991}, \emph{360}, 362 -- 396\relax
\mciteBstWouldAddEndPuncttrue
\mciteSetBstMidEndSepPunct{\mcitedefaultmidpunct}
{\mcitedefaultendpunct}{\mcitedefaultseppunct}\relax
\EndOfBibitem
\bibitem[Read and Green(2000)Read, and Green]{Read-Green}
Read,~N.; Green,~D. Paired states of fermions in two dimensions with breaking
  of parity and time-reversal symmetries and the fractional quantum Hall
  effect. \emph{Phys. Rev. B} \textbf{2000}, \emph{61}, 10267--10297\relax
\mciteBstWouldAddEndPuncttrue
\mciteSetBstMidEndSepPunct{\mcitedefaultmidpunct}
{\mcitedefaultendpunct}{\mcitedefaultseppunct}\relax
\EndOfBibitem
\bibitem[Dukelsky \latin{et~al.}(2016)Dukelsky, Pittel, and Esebbag]{Pittel}
Dukelsky,~J.; Pittel,~S.; Esebbag,~C. Structure of the number-projected BCS
  wave function. \emph{Phys. Rev. C} \textbf{2016}, \emph{93}, 034313\relax
\mciteBstWouldAddEndPuncttrue
\mciteSetBstMidEndSepPunct{\mcitedefaultmidpunct}
{\mcitedefaultendpunct}{\mcitedefaultseppunct}\relax
\EndOfBibitem
\bibitem[Richardson(1963)]{Rich63}
Richardson,~R. A restricted class of exact eigenstates of the pairing-force
  Hamiltonian. \emph{Physics Letters} \textbf{1963}, \emph{3}, 277 -- 279\relax
\mciteBstWouldAddEndPuncttrue
\mciteSetBstMidEndSepPunct{\mcitedefaultmidpunct}
{\mcitedefaultendpunct}{\mcitedefaultseppunct}\relax
\EndOfBibitem
\bibitem[von Delft \latin{et~al.}(1996)von Delft, Zaikin, Golubev, and
  Tichy]{Delft96}
von Delft,~J.; Zaikin,~A.~D.; Golubev,~D.~S.; Tichy,~W. Parity-Affected
  Superconductivity in Ultrasmall Metallic Grains. \emph{Phys. Rev. Lett.}
  \textbf{1996}, \emph{77}, 3189--3192\relax
\mciteBstWouldAddEndPuncttrue
\mciteSetBstMidEndSepPunct{\mcitedefaultmidpunct}
{\mcitedefaultendpunct}{\mcitedefaultseppunct}\relax
\EndOfBibitem
\bibitem[Ortiz and Dukelsky(2005)Ortiz, and Dukelsky]{crossover}
Ortiz,~G.; Dukelsky,~J. BCS-to-BEC crossover from the exact BCS solution.
  \emph{Phys. Rev. A} \textbf{2005}, \emph{72}, 043611\relax
\mciteBstWouldAddEndPuncttrue
\mciteSetBstMidEndSepPunct{\mcitedefaultmidpunct}
{\mcitedefaultendpunct}{\mcitedefaultseppunct}\relax
\EndOfBibitem
\bibitem[Dukelsky and Sierra(1999)Dukelsky, and Sierra]{Sierra1999}
Dukelsky,~J.; Sierra,~G. Density Matrix Renormalization Group Study of
  Ultrasmall Superconducting Grains. \emph{Phys. Rev. Lett.} \textbf{1999},
  \emph{83}, 172--175\relax
\mciteBstWouldAddEndPuncttrue
\mciteSetBstMidEndSepPunct{\mcitedefaultmidpunct}
{\mcitedefaultendpunct}{\mcitedefaultseppunct}\relax
\EndOfBibitem
\bibitem[Degroote \latin{et~al.}(2016)Degroote, Henderson, Zhao, Dukelsky, and
  Scuseria]{Scuseria}
Degroote,~M.; Henderson,~T.~M.; Zhao,~J.; Dukelsky,~J.; Scuseria,~G.~E.
  Polynomial similarity transformation theory: A smooth interpolation between
  coupled cluster doubles and projected BCS applied to the reduced BCS
  Hamiltonian. \emph{Phys. Rev. B} \textbf{2016}, \emph{93}, 125124\relax
\mciteBstWouldAddEndPuncttrue
\mciteSetBstMidEndSepPunct{\mcitedefaultmidpunct}
{\mcitedefaultendpunct}{\mcitedefaultseppunct}\relax
\EndOfBibitem
\bibitem[Ripoche \latin{et~al.}(2017)Ripoche, Lacroix, Gambacurta, Ebran, and
  Duguet]{Duguet}
Ripoche,~J.; Lacroix,~D.; Gambacurta,~D.; Ebran,~J.-P.; Duguet,~T. Combining
  symmetry breaking and restoration with configuration interaction: A highly
  accurate many-body scheme applied to the pairing Hamiltonian. \emph{Phys.
  Rev. C} \textbf{2017}, \emph{95}, 014326\relax
\mciteBstWouldAddEndPuncttrue
\mciteSetBstMidEndSepPunct{\mcitedefaultmidpunct}
{\mcitedefaultendpunct}{\mcitedefaultseppunct}\relax
\EndOfBibitem
\end{mcitethebibliography}

\end{document}